\documentclass[12pt,a4paper]{article}
\usepackage{epsfig,graphicx,axodraw}

\newcommand{\newc}{\newcommand}
\newc\eg{{\it {e.g.}}}

\newcommand\stau{\widetilde{\tau}}

\newcommand\neut{\tilde \chi}

\newc{\cachigamma}{C_{a\neut\gamma}}
\newc{\caww}{C_{aWW}}
\newc{\cayy}{C_{aYY}}
\newc{\sthw}{\sin\theta_W}
\newc{\cthw}{\cos\theta_W}
\newc{\bino}{\widetilde B}
\newc{\wino}{\widetilde W_3}
\newc{\higgsinob}{{\widetilde H}^0_b}
\newc{\higgsinot}{{\widetilde H}^0_t}
\newc{\abund}{\Omega h^2}
\newc{\abundchi}{\Omega_\neut h^2}
\newc{\rhocrit}{\rho_{crit}}
\newc{\rhochi}{\rho_{\neut}}
\newc{\mplanck}{M_{\rm P}}
\newc{\xf}{x_f}
\newc{\jxf}{J({\xf})}
\newc{\VEV}[1]{\langle #1 \rangle}
\newc{\nl}{\cos \theta_{\tilde t}}
\newc{\nr}{\sin \theta_{\tilde t}}
\newc{\mgluino}{m_{\tilde g}}

\newc\gbar{{\overline{g}}}

\newc{\ra}{\rightarrow}
\newc{\beq}{\begin{equation}}
\newc{\eeq}{\end{equation}}
\newc{\bea}{\begin{eqnarray}}
\newc{\eea}{\end{eqnarray}}
\renewcommand{\[}{\left[}

\newc{\nspin}{n_{\rm spin}}
\newc{\nflavor}{n_{\rm F}}
\newc\vrel{v_{\rm rel}}

\newc{\naxino}{n_{\tilde a}}
\newc{\ngamma}{n_\gamma}
\newc{\ychi}{Y_{\chi}}                  \newc{\yeqchi}{Y^{\rm EQ}_{\chi}}
\newc{\yaxino}{Y_{\tilde a}}
\newc{\yeqaxino}{Y^{\rm EQ}_{\tilde a}}
\newc{\ythaxino}{Y^{\rm TP}_{\tilde a}}
\newc{\ynthaxino}{Y^{\rm NTP}_{\tilde a}}
\newc{\yascat}{Y^{\rm scat}_{i,j}}      \newc{\yadec}{Y^{\rm dec}_{i}}
\newc{\gstar}{g_\ast}           \newc{\gsstar}{g_{s\ast}}
\newcommand\lsim{\mathrel{\rlap{\lower4pt\hbox{\hskip1pt$\sim$}}
    \raise1pt\hbox{$<$}}}
\newcommand\gsim{\mathrel{\rlap{\lower4pt\hbox{\hskip1pt$\sim$}}
    \raise1pt\hbox{$>$}}}

       \def\pslash{\not{\hbox{\kern-2.3pt $p$}}}
       \def\kslash{\not{\hbox{\kern-2.3pt $k$}}}
       \def\qslash{\not{\hbox{\kern-2.3pt $q$}}}
       \def\ddslash{\not{\hbox{\kern-2.3pt $d$}}}

\def\apj#1#2#3{Astrophys.\ J.\ {\bf #1} (#2) #3}

\begin{document}

\thispagestyle{empty}

\vspace*{-12mm}
\begin{flushright}
\today \hfill DESY 08-035 \\
 LPSC 08-40 \\ 
 MIT-CTP-3936 \\ 
NSF-KITP-08-49 \\
 LYCEN 2008-10 \\
\end{flushright}

\vspace*{2cm}

\begin{center}

{\LARGE The number density of a charged relic}\\[12mm]

{\large Carola F. Berger$^{1,2}$, Laura Covi$^{3}$, Sabine Kraml$^{4}$ \\[2mm]
           and  Federica Palorini$^{5}$ }\\[12mm]

{\it
$^{(1)}$ Center for Theoretical Physics, Massachusetts Institute of Technology,\\
              Cambridge, MA 02139, USA \\
$^{(2)}$ Kavli Institute for Theoretical Physics, University of California,\\
              Santa Barbara, CA 93106-4030, USA \\
$^{(3)}$ DESY Theory Group, Notkestrasse 85, D-22603 Hamburg, Germany \\
$^{(4)}$ Laboratoire de Physique Subatomique et de Cosmologie,
              UJF Grenoble 1, CNRS/IN2P3,
              53 Avenue des Martyrs, F-38026 Grenoble, France\\
$^{(5)}$ IPNL, Universit\'e de Lyon, Universit\'e Lyon 1, CNRS/IN2P3, \\ 4 rue E. Fermi, F-69622 Villeurbanne, France

}

\vspace*{12mm}

\begin{abstract}
\noindent We investigate scenarios in which a charged, long-lived
scalar particle decouples from the primordial plasma in the Early
Universe. We compute the number density at time of freeze-out
considering both the cases of abelian and non-abelian interactions
and including the effect of Sommerfeld enhancement at low initial
velocity. We also discuss as extreme case the maximal cross
section that fulfils the unitarity bound. We then compare these
number densities to the exotic nuclei searches for stable relics
and to the BBN bounds on unstable relics and draw conclusions for
the cases of a stau or stop NLSP in supersymmetric models with a
gravitino or axino LSP.
\end{abstract}

\end{center}

\vspace*{6mm}

\newpage
\tableofcontents

\section{Introduction}

The early Universe may have been populated by many exotic particles that,
especially if charged, should have easily been in thermal equilibrium.
No charged relic seems to have survived to the present day.
In fact there are very strong upper bounds on the density of
electromagnetically 
and/or colour charged particles with masses below 10--100 TeV from
extensive searches for exotic nuclei \cite{ch-relics}.
The standard lore is therefore that only neutral relics may have survived
until today.

However, it is possible that some unstable but very long-lived
charged particle froze-out from thermal equilibrium and decayed
much later to a neutral one. A typical example of this kind in
supersymmetric models with R-parity conservation is the
next-to-lightest supersymmetric particle~(NLSP) if the LSP and
Cold Dark Matter is very weakly interacting like the 
axino~\cite{axinolsp,crs02,axinolsp2} or the 
gravitino~\cite{gravitinolsp, neutralNLSP}. 
Recently, such candidates have attracted a lot of attention, and indeed 
the signal of a charged metastable NLSP at colliders would be
spectacular~\cite{stauatcolliders,Fairbairn:2006gg}.

In general, strong bounds on the number density of any metastable relic 
with lifetime of about 1\,s or longer are provided by Big Bang 
Nucleosynthesis (BBN)~\cite{BBNrev}. 
They come from two classes of processes: on one hand injection of very 
energetic photons or hadrons from decays during or after BBN adds 
an additional non-thermal component to the plasma and can modify the 
abundances of the light elements~\cite{neutBBN}; on the other hand, if 
the relic particle is electromagnetically charged, bound states with 
nuclei may arise that strongly enhance some of the nuclear rates and 
allow for catalysed production of e.g.\ $^6Li$~\cite{CBBN}.
The bounds of the first type are very tight for lifetimes of the order 
of $10^4$\,s and exclude, for instance, a neutralino NLSP with a
gravitino LSP in the CMSSM~\cite{neutralNLSP}.
An electrically charged NLSP like the $\tilde\tau$ can instead escape 
the first class of constraints in part of the parameter space, but it 
is excluded for long lifetimes by bound state effects~\cite{stauNLSP}.
In the axino LSP case, the NLSP has a shorter lifetime;
the BBN bounds are hence much weaker and both, neutralino and stau, NLSP
are still allowed \cite{axinolsp}.

In this paper, we investigate the most general case of a scalar charged thermal relic.
We compute the number density and compare it to the bounds on exotic nuclei for
stable particles and the BBN constraints for unstable ones.
Similar studies have been carried out model-independently many years 
ago~\cite{wolfram, Nardi:1990ku, unitarity} for stable relics and 
we will update and 
improve these computations.\footnote{Recently the case of general 
EW charged relics as DM was also considered in full 
detail~\cite{Cirelli07}.}
We mostly consider the role of the gauge interaction for two main reasons:
i) the annihilation into gauge bosons is often the dominant channel for a charged particle and
ii) it depends only on very few parameters, just the mass of the particle and its charge or
representation. 
It is also enhanced by the Sommerfeld effect~\cite{Sommerfeld},  analogous
to heavy quark production at threshold,  which has previously been considered 
for dark matter annihilations in 
\cite{gluino-splitsusy, Hisano:2006nn, Cirelli07, 
Freitas07, DMsommerfeld} and 
recently also in the
context of leptogenesis in \cite{Strumia:2008cf}.
We discuss this Sommerfeld enhancement for the general abelian and 
non-abelian cases.
Moreover, we compare the cross sections with the unitarity bound and
update the unitarity limit on the mass of a stable relic.

Our main goal is to determine if it is at all possible to evade {\it completely} either
the exotic nuclei bounds or the BBN ones and how strongly the particle has to interact
in this case. We then apply our findings to the Minimal Supersymmetric Standard
Model and discuss in more detail the cases of the stau and stop NLSP.

The paper is organised as follows. In Section~2, we briefly review
the computation of the number density from thermal freeze-out. The
formulae for the annihilation cross section of a charged particle
into gauge bosons are given in Section~3. Here we discuss abelian
and non-abelian cases, the Sommerfeld enhancement and the unitarity
cross section. Moreover, we compare the thermal averages with the first
order in velocity expansion. The resulting relic density is
discussed in Section~4. In Section~5, we review the constraints on
stable and unstable relics. These are then applied in Section~6 to
the concrete examples of relic staus and stops. Section~7 finally
contains our conclusions. Details on the computation of the
annihilation cross section and the case of massive gauge bosons
are given in the Appendices A and B.

\section{Number density of a thermal relic}

The number density of a stable or quasi-stable thermal relic is
determined by its annihilation cross section. In fact the
number density of a particle in a thermal bath and an expanding
Universe is described by the Boltzmann equation \cite{kt91,gg91}:
\beq
\dot n_X + 3 H n_X = \int {dp_X^3\over (2\pi)^3 2 E_X}\; {\cal C} [f_X]
\label{Boltzmann-1}
\eeq
where the dot indicates the time derivative,
${\cal C}$ denotes the collision integral of all
processes that change the particle number and $f_X$ is
the phase-space density for the particle $X$.
For a particle with a conserved parity, like R-parity,
the lowest order processes to be considered in the
collision integral are just two particle scatterings,
i.e. annihilations and coannihilations.
If there is a lighter particle carrying the conserved
parity number, ${\cal C}$ includes also the decay into
this lighter state,
but we will assume that such a decay rate is so small
it can be neglected at the time of freeze-out and becomes
effective only much later. Then we have effectively a two step
process and we can treat freeze-out and decay separately.
This is a general feature if the decay takes place via
a non-renormalisable interaction
and is suppressed by an intermediate or even the Planck scale
(see e.g. the axino \cite{axinolsp, axinolsp2} and gravitino 
cases \cite{gravitinolsp, neutralNLSP}).

Taking into account only the annihilation of particle and
antiparticle, we can write the
collision integral as \cite{gg91}
\beq
{\cal C}[f_X] =
 - \int {dp_{\bar X}^3\over (2\pi)^3 2 E_{\bar X}}
\left(f_X f_{\bar X} - f_X^{eq} f_{\bar X}^{eq} \right)
4 \sqrt{(p_X \cdot p_{\bar X})- m^4_X}
\;\sigma_{ann}
\label{collision-int}
\eeq
where $\sigma_{ann} $ denotes the unpolarised annihilation cross section
of an $X \bar X$ pair summed over initial and final states.
We are here assuming that CP is conserved and no asymmetry exists
between $n_X$ and $n_{\bar X}$. Note that the production cross section
is taken into account by the term proportional to $  f_X^{eq} f_{\bar X}^{eq} $
since we are assuming that the products of the annihilation are much
lighter than $X$ and are still in thermal equilibrium.

In this paper we will consider charged relics and concentrate therefore
on the annihilation into gauge bosons, which is the dominant channel
in most of parameter space and does depend only on the mass and charge of
the relic. Note that adding more channels only increases the
cross section and reduces the
relic particle number density further. Instead, the inclusion of
coannihilations for a charged particle does not always reduce the 
number density as discussed in \cite{ahs00}.

We can rewrite eq.~(\ref{Boltzmann-1}) by changing variable to
$Y_X = n_X/s$, where $s(T) = g_S {2\pi^2\over 45} T^3 $ is the entropy
density, so that the dilution due to the expansion of the universe
cancels out in the ratio as long as entropy is conserved.
It is also convenient to replace the time variable with
$x = {m_X\over T}$, thanks to the relation $ dt = {dx \over (xH)} $.
We thus obtain
\bea
{dY_X \over dx} &=& - { x s(x) \over H(x) m_X^2}
\langle \sigma v \rangle_x
\left( Y_X^2 - Y_{eq}^2 \right)\\
&=& - {2\pi g_S\over 15} \left({10\over g_{\rho}} \right)^{1/2}
{M_P \over m_X} \langle \sigma v \rangle_x
\left( Y_X^2 - Y_{eq}^2 \right)\; .
\label{Boltzmann-2}
\eea
Here we have used $H^2 = {\pi^2\over 90} g_{\rho}
{T^4\over M_P^2} $, for $M_P = 2.43 \times 10^{18} $ GeV,
valid during the radiation dominated era.
Moreover, we define the thermally averaged cross section
as\footnote{Note that our definition differs from the one in~\cite{gg91}
by a factor $ m_X^2/x^2 $ since we prefer to work with a dimensionless
quantity and to absorb here all the dependence on $x$.}
\beq
 \langle \sigma v \rangle_x = { 1 \over4 x^4 K_2^2(x) }
\;\; \int_{2x}^{\infty} dz z^2 \tilde \sigma \left({x\over z}\right) K_1(z)
\label{sigmatermica}
\eeq
where $K_i (z)$ are the modified Bessel functions of order $i$,
characteristic of  Maxwell-Boltzmann statistics (we are
assuming that we can approximate Bose-Einstein statistics
with Maxwell-Boltzmann statistics).
In this expression the rescaled cross section
$\tilde \sigma $ is given by the annihilation
cross section  {\it averaged} over initial and  {\it summed} over final
states and multiplied by a factor proportional to the squared M\o ller velocity,
\beq
\tilde \sigma \left( {m_X\over \sqrt{s}} \right)
= (s - 4m_X^2) \sigma (m_X,s)\; .
\eeq
Note that in the centre-of-mass system the M\o ller velocity is
equal to the relative velocity between the annihilating particles
 and given by
\beq
v_{\rm M\o l} = 2\beta = 2 \sqrt{1 - \frac{4m_X^2}{s}}\; .
\eeq
The rescaled cross section $ \tilde \sigma $ defined above is dimensionless 
and function only of  $x/z = m_X/\sqrt{s}$ (or $\beta $) for the case 
of annihilation into massless gauge bosons and it always vanishes at threshold.
Then it is easy to see that since we integrate in both $x, z$, 
the main dependence on the charged relic mass is contained in the prefactor 
in eq.~(\ref{Boltzmann-2}) and can be reabsorbed in a rescaling of  
$ Y_X \rightarrow Y_X/m_X $. For this reason we obtain nearly exactly 
$ Y_X \propto m_X $ if there is no other mass scale involved.
Note that in principle a much weaker logarithmic dependence on $m_X$ is 
present in the value of the freeze-out temperature, when $Y $ begins 
to deviate from  $ Y_{eq} $.

We are here computing the yield of the particle $X$ and to obtain the yield of particle
and antiparticle we multiply by a factor of 2 or divide the cross section by 1/2,
since we are assuming $n_X = n_{\bar X}$. Also note that, contrary to intuition, for a
particle with internal degrees of freedom like a coloured state,  the total yield is the
solution of the Boltzmann equation (\ref{Boltzmann-2}) with the cross section averaged
over the initial states. Instead the yield per degree of freedom is obtained from the
cross section averaged over $X$, but summed over $\bar X$~\footnote{
In fact any rescaling of the cross section by a factor $p$ due to a different counting of the
degrees of freedom can be absorbed into a rescaling $1/p$ of the yield(s). }.
The presence of many degrees of freedom in the initial state has then the effect of partially
compensating the large cross section coming from the multiplicity of the final states.

\section{Annihilation cross section for a charged particle into gauge bosons}

\subsection{Abelian case}

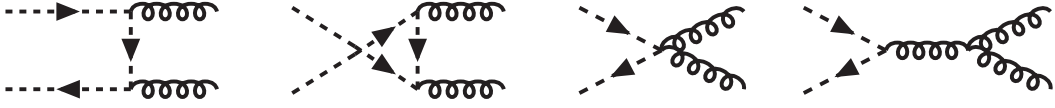
\begin{figure}
\unitlength=1.0 pt
\SetScale{1.46}
\SetWidth{1.0}
\begin{center}
\begin{picture}(80,30)(0,0)
\DashArrowLine(0,20)(32,20){2}
\DashArrowLine(32,20)(32,0){2}
\DashArrowLine(32,0)(0,0){2}
\Gluon(32,20)(54,20){2}{4}
\Gluon(32,0)(54,0){2}{4}
\end{picture}
\qquad
\begin{picture}(80,30)(0,0)
\DashLine(0,21)(16,11){2}
\DashArrowLine(16,11)(32,0){2}
\DashArrowLine(32,20)(32,0){2}
\DashLine(0,-1)(16,9){2}
\DashArrowLine(16,9)(32,20){2}
\Gluon(32,20)(54,20){2}{4}
\Gluon(32,0)(54,0){2}{4}
\end{picture} \qquad
\begin{picture}(80,30)(0,0)
\DashArrowLine(0,21)(21,10){2}
\DashArrowLine(21,10)(0,-1){2}
\Gluon(21,10)(42,20){2}{4}
\Gluon(21,10)(42,0){2}{4}
\end{picture}
\begin{picture}(80,30)(0,0)
\DashArrowLine(0,21)(21,10){2}
\DashArrowLine(21,10)(0,-1){2}
\Gluon(21,10)(42,10){2}{4}
\Gluon(42,10)(63,20){2}{4}
\Gluon(42,10)(63,0){2}{4}
\end{picture}
\end{center}
\caption{\label{FD-1}
Feynman diagrams for the annihilation into gauge bosons, here for the case of gluons.
In the abelian case, there is no 3-gauge-boson vertex, so the last diagram is absent.}
\end{figure}

For an abelian gauge symmetry, there are only  three Feynman diagrams contributing to the
annihilation cross section, analogous to those shown in Fig.~\ref{FD-1}: the t- and u-channel
exchange of the scalar particle itself, and the 4-boson vertex.
The amplitude is symmetric in the exchange of the gauge bosons and for a particle of charge
$e_X g_1 $ it is given by
\beq
{\cal A}^{\mu\nu} =
i g_1^2 e_X^2 \left[
{(2 p_1 - p_3)^{\mu} (2 p_2 - p_4)^{\nu} \over t-m_X^2}
+ {(2 p_1 - p_4)^{\nu} (2 p_2 - p_3)^{\mu} \over u-m_X^2}
+ 2 g^{\mu\nu} \right]\; .
\eeq
The cross section is a function of the mass and charge of the relic:
\bea
\sigma_{\mathrm ab} (m_X, s) &=& {4\pi \alpha_1^2 e_X^4 \over s - 4m_X^2}
\left[
\sqrt{1-{4m_X^2\over s} } \left( 1 + {4 m_X^2 \over s}\right)
\right.\nonumber\\
& & \left. +\; {4m_X^2\over s} \left(1-{2m_X^2\over s}\right)
\log \left({1-\sqrt{1-{4m_X^2\over s}}\over
1+ \sqrt{1-{4m_X^2\over s}}} \right)
\right]
\label{sigmaab}
\eea
where $\alpha_1 = g_1^2/(4\pi ) $ is the gauge coupling; note that
a symmetry  factor $1/2$ has to be added due to the symmetric final state
of identical particles.
For the rescaled cross section this gives
\bea
\tilde \sigma_{\mathrm ab} (\beta) &=& 8\pi \alpha_1^2 e_X^4 \beta
\left[
1 - \frac{1}{2} \beta^2
+ \frac{1- \beta^4}{ 4\beta}
\log \left({1-\beta \over
1+\beta} \right)
\right]\,,
\label{eq-sigma-abelian}
\eea
which is a function only of $\beta = \sqrt{1 - 4 m_X^2/s} $ and the charge of the particle.

\subsection{Non-abelian case}

%

The computation for the annihilation into non-abelian gauge bosons is slightly more involved,
since there is an additional contribution from the Feynman diagram with a gauge boson
in the s-channel and the 3-gauge-boson vertex.
The amplitude can be divided into a symmetric and an antisymmetric piece in the group indices.
The symmetric one is analogous to the abelian case:
\bea
{\cal A}^{\mu\nu}_{sym} &=&
i {g_N^2 \over 2} \left\{T^a, T^b \right\}_{ji}
\left[{(2 p_1 - p_3)^{\mu} (2 p_2 - p_4)^{\nu} \over t-m_X^2}
\right.\nonumber\\
& & \left.
+\; {(2 p_1 - p_4)^{\nu} (2 p_2 - p_3)^{\mu} \over u-m_X^2}
+ 2 g^{\mu\nu} \right]\, ,
\eea
while the antisymmetric part is given by
\bea
{\cal A}^{\mu\nu}_{asym} &=&
- i {g_N^2\over 2} \left[ T^a, T^b \right]_{ji}
\left[
{(2 p_1 - p_3)^{\mu} (2 p_2 - p_4)^{\nu} \over t-m_X^2}
- {(2 p_1 - p_4)^{\nu} (2 p_2 - p_3)^{\mu} \over u-m_X^2}
\right. \nonumber\\
& & \left.
+\; 2 { g^{\mu\nu} (t-u) - (2p_4+p_3)^{\mu} (p_1-p_2)^{\nu} +
(p_1-p_2)^{\mu} (2p_3+p_4)^{\nu} \over s}
\right]\, .
\eea
The two contributions do not interfere due to the different symmetry, so we have for the
amplitude squared,  summing only over physical polarisations of the final gauge bosons:
\bea
|{\cal M}|^2 &=& 4 g_N^4 \left\{
\Big|\left\{T^a, T^b \right\}_{ji}\Big|^2
\[{1\over 2} + { 2 m_X^4 \over (t-m_X^2)^2} + {2m_X^2 \over t-m_X^2}
\left( 1- {2m_X^2\over s} \right)\right]
\right. \nonumber\\
& & \left. +\; \Big|\left[T^a, T^b \right]_{ji}\Big|^2
\left[ {1\over 2} {(s + 2 (t-m_X^2))^2 \over s^2}
+ {4 m_X^2 \over s}
+ { 2 m_X^4 \over (t-m_X^2)^2}
\right.\right. \nonumber\\
& & \left. \left.
+\; {2m_X^2 \over t-m_X^2}
\left( 1+ {2m_X^2\over s} \right)\right] \right\} \, .
\eea

Then the sum over all final and initial states for a scalar in the
fundamental representation $T^a$ of the gauge group $SU(N)$,
normalised such that $ {\mathrm Tr} (T^a T^b) = \delta^{ab}/2 $,
can be obtained from the usual group invariants:
\bea
\sum_{j,i,a,b} \frac{1}{2} \Big|\left\{T^a, T^b \right\}_{ji}\Big|^2 
&=& 
\sum_{a,b} \frac{1}{2} \left (\frac{1}{N} \delta^{ab} 
+ \frac{1}{2} \sum_c | d_{abc} |^2 \right) \nonumber \\
&=& C_F (N) \left(1 + \frac{1}{2} (C_A^2(N)-4) \right) =
\frac{(N^2-1)(N^2-2)}{ 4 N}\; ,
\eea
where we have separated the singlet and adjoint contributions
to the symmetric part for later convenience, included a factor $1/2$
for identical particles in the final states and used
the Casimir invariants for the fundamental and adjoint representations,
 $ C_F(N) = \frac{N^2-1}{2N}, C_A(N) = N $. Note that the ratio of 
the singlet to adjoint contributions is given simply by
$ \frac{2}{N^2-4} $.
The antisymmetric channel instead gives
\beq
\sum_{j,i,a,b} \frac{1}{2} \Big|\left[T^a, T^b \right]_{ji}\Big|^2 = 
\frac{N^2-1}{4} C_A(N) =
\frac{N (N^2-1)}{ 4} \; .
\eeq

Finally we obtain for the cross section averaged over initial states:
\bea
\label{sigmaN}
\sigma_{nab} (m_X, s) &=&
{\pi \alpha_N^2 \over s - 4m_X^2} {(N^2-1)^2\over N^3} \times \nonumber\\
&& \left[
\sqrt{1-{4m_X^2\over s} }
\left(1+ {4m_X^2\over s}
- {N^2\over 3 (N^2-1)} \left( 1 - {10 m_X^2\over s}\right) \right)
\right.\nonumber\\
& & \left. \;+\;  4 {m_X^2\over s}
\left(1 + {2\over N^2-1} {m_X^2\over s}\right)
\log \left({1-\sqrt{1-{4m_X^2\over s}}\over
1+ \sqrt{1-{4m_X^2\over s}}} \right)
\right]\, .
\eea
This result coincides for $N=3$ with that reported in \cite{bhsz96}
for the Born cross section of a pair of gluons into squarks,
allowing for the exchange of the initial and final state.
%

Then the rescaled cross section for $SU(N)$ is
\bea
\tilde\sigma_{nab} (\beta ) &=&
2 \pi \alpha_N^2 \frac{(N^2-1)^2 }{N^3} \beta
\left[
1 + \frac{N^2}{4(N^2 -1)} - \frac{\beta^2}{2}
\left(1+ \frac{ 5 N^2 }{ 6 (N^2-1)} \right)
\right.\nonumber\\
& & \left.
+\; \frac{1-\beta^2}{2\beta}
\left(1 +  \frac{1}{2 (N^2-1)} - \frac{\beta^2}{2 (N^2-1)} \right)
\log \left(\frac{1-\beta}{ 1+ \beta} \right)
\right]\, .
\label{eq-sigma-nonabelian}
\eea
Note that the contribution of order $ \beta $ in the expression above
in the limit  $ \beta \rightarrow 0 $ is due to the symmetric part of 
the matrix element and that the antisymmetric piece instead vanishes
at that order. Therefore the symmetric part of the cross section
dominates at threshold.
  
So we see that for a non-abelian interaction the cross section is larger 
than for the abelian case, not only due to the possibly larger coupling 
$\alpha_N $, but also due to the opening of an antisymmetric channel and 
of course to the multiplicity of the final states. In fact for large $N$ 
the averaged cross section increases as $N$ and therefore the yield 
decreases as $1/N$.

\subsection{Annihilation into $SU(N)$ gauge boson and photon}

The annihilation cross section into gluon and photon is just the same as
the abelian one, but with a different vertex for the gluon.
Then considering a particle of electromagnetic charge $e_X g_1$, in the
representation $T^a$ of the gauge group $SU(N) $ with coupling $g_N$,
annihilating with its own antiparticle, the amplitude
is given by~\footnote{Strictly speaking, in this case the final state particles are
different and therefore there are no independent t- and u-channels,
but we can still write the amplitude to be symmetric in t and u
in order to make direct contact with the previous results.}
\beq
{\cal A}^{\mu\nu} =
i g_1 e_X  g_N T^a_{ji} \left[
{(2 p_1 - p_3)^{\mu} (2 p_2 - p_4)^{\nu} \over t-m_X^2}
+ {(2 p_1 - p_4)^{\nu} (2 p_2 - p_3)^{\mu} \over u-m_X^2}
+ 2 g^{\mu\nu} \right]\; .
\eeq
From this we easily obtain the cross section as:
\bea
\sigma_{\mathrm 1N} (m_X, s) &=&
{8\pi \alpha_1 \alpha_N e_X^2 \over s - 4m_X^2}\,   |T^a_{ji} |^2
\left[
\sqrt{1-{4m_X^2\over s} } \left( 1 + {4 m_X^2 \over s}\right)
\right.\nonumber\\
& & \left. +\; {4m_X^2\over s} \left(1-{2m_X^2\over s}\right)
\log \left({1-\sqrt{1-{4m_X^2\over s}}\over
1+ \sqrt{1-{4m_X^2\over s}}} \right)
\right]\,,
\eea
where $\alpha_{1,N} $ are the gauge couplings and the symmetry
factor $1/2$ in this case is absent since the final particles are
not identical.

\noindent
Averaging over the initial and summing over the final state, we have
\beq
\frac{1}{N} \sum_{j,i,a} T^a_{ji} T^a_{ij}
= \frac{1}{N} \sum_{j,i} C_F(N) \delta_{ij}
= \frac{N^2 - 1}{2N}
\label{N-ave2}
\eeq
for the fundamental representation. 
This gives for the rescaled cross section
\bea
\tilde \sigma_{\mathrm 1N} (\beta) &=& 8\pi \alpha_1
 \alpha_N e_X^2
 \frac{N^2 - 1}{N}
 \beta \left[
1 - \frac{1}{2} \beta^2
+ \frac{1- \beta^4}{ 4\beta}
\log \left({1-\beta \over
1+\beta} \right)
\right]
\eea
which is a factor $ (N^2-1) \alpha_N/ ( N \alpha_1 e_X^2) $
larger than the pure $U(1)$ contribution.
Again the cross section increases as $N$ for large $N$.

\subsection{Annihilation into physical Z and $SU(N)$ gauge boson/photon}

The annihilation cross section into massive Z and photon/SU(N) gauge boson
has the same form as the abelian one. We consider here a particle with Z-coupling
$ g_1 e_Z $, in the representation $T^a$ of the gauge group $SU(N) $
with coupling $g_N$, annihilating with its own antiparticle and we obtain
\beq
{\cal A}^{\mu\nu} =
i g_1 e_Z  g_N T^a_{ji} \left[
{(2 p_1 - p_3)^{\mu} (2 p_2 - p_4)^{\nu} \over t-m_X^2}
+ {(2 p_1 - p_4)^{\nu} (2 p_2 - p_3)^{\mu} \over u-m_X^2}
+ 2 g^{\mu\nu} \right]\; ,
\eeq
where $p_4$ is the Z boson momentum obeying $ p_4^2 = M_Z^2 $;
the annihilation into photon and Z is easily read off by taking
just $ g_N  T^a_{ji} \rightarrow g_1' e_X $.
Then we easily obtain the cross section as:
\bea
\sigma_{\mathrm ZN} (m_X, M_Z, s) &=&
{8\pi \alpha_1 \alpha_N e_Z^2 \over s - 4m_X^2}\,   |T^a_{ji} |^2
\left[
\sqrt{1-{4m_X^2\over s} } \left( 1 - \frac{M_Z^2}{s}
+ \frac {4 (m_X^2-M_Z^2)}{ s - M_Z^2}\right)
\right.\nonumber\\
& & \left. +\; {4m_X^2\over s}
\left(1- \frac{ 5 M_Z^2}{ 8 m_X^2} - \frac{4m_X^2- 3 M_Z^2}{ 2 (s-M_Z^2)}\right)
\log \left({1-\sqrt{1-{4m_X^2\over s}}\over
1+ \sqrt{1-{4m_X^2\over s}}} \right)
\right]\,,
\eea
where $\alpha_{1,N} $ are the gauge couplings.

\noindent
Averaging over the initial and summing over the final state as in
eq.~(\ref{N-ave2}), we have for the rescaled cross section
\bea
\tilde \sigma_{\mathrm 1N} (\beta, a_Z ) &=& 8\pi \alpha_1
 \alpha_N e_Z^2
 \frac{N^2 - 1}{2N}
 \beta \left[1 - a_Z  ( 1- \beta^2)
 + \frac{(1 - 4 a_Z) (1-\beta^2)}{ 1- a_Z (1-\beta^2)}
 \right.\nonumber\\
& & \left.
 + \frac{1- \beta^2}{ \beta}
 \left(    1 - \frac{5}{2} a_Z -  (1-\beta^2) \frac{1- 3 a_Z}{2 - 2 a_Z (1-\beta^2)}
   \right)
\log \left({1-\beta \over
1+\beta} \right)
\right]\; ,
\eea
where $a_Z = M_Z^2/m_X^2 $.
Note that the cross section for annihilation into photon and Z, is given by
the substitution $ \alpha_N \frac{N^2 - 1}{2N}  \rightarrow \alpha_1' e_X^2 $.
For the specific case of the right-handed stau (stop), the coupling with the Z boson
and photon are respectively given by
$ e_Z^2 \alpha_1 =  \alpha_{em} \tan^2\theta_W $
($ e_Z^2\alpha_1 = 4/9 \,\alpha_{em}   \tan^2\theta_W   $) and
$ e_X^2 \alpha_1' = \alpha_{em}$ ($ e_X ^2 \alpha_1' =  4/9\, \alpha_{em} $),
where $\theta_W$ is the Weinberg angle.

\subsection{Annihilation into massless EW gauge bosons}

The cross section for annihilation into massless $SU(2)_L $ gauge bosons
can be obtained directly from the general formula for the non-abelian case.
One has to take into account, however, that in this case the scalar $SU(2)_L$
doublet is not degenerate in mass and that the initial particles
can be a mixture of left- and right-chiral states.
We neglect here the effects of EW symmetry breaking; the results are hence
applicable for the case of a heavy relic that decouples before EW symmetry
breaking takes place.

Considering the scalar relic to be
$X =  X_L \cos\theta +  X_R \sin\theta$ and denoting with
$m_{X'} $ the mass of its left-handed doublet partner, which is sufficiently
larger than $m_X $ to neglect coannihilations, we obtain
for the annihilation cross section into $ W^{1,2} $ gauge
bosons: 
\bea
& &  \hspace*{-4mm} \sigma_{W2} (s, m_X, m_{X'})  = \nonumber\\[3mm]
& & \frac{2\pi\alpha_2^2 \cos^4\theta}{ s - 4m_X^2}
\left[ \sqrt{1 - \frac{4 m_X^2}{s}}
\left(
\frac{2}{3} + \frac{13}{3} \frac{ m_X^2}{s}
- \frac{m_{X'}^2}{s}
+ \frac{(m_X^2+m_{X'}^2)^2 }{s m_{X'}^2 + (m_{X'}^2-m_X^2)^2}
\right)
\right.\\
& & \left.
+\;  2 \left( \frac{m_{X'}^2 + m_X^2}{s}
- \frac{(m_{X'}^2 - m_X^2)^2 }{ 2 s^2}\right)
\log \left( \frac{s + 2 (m_{X'}^2 - m_X^2) - \sqrt{s(s- 4m_X^2)} }{
s + 2 (m_{X'}^2 - m_X^2) + \sqrt{s(s- 4m_X^2)}} \right)
\right] \nonumber \, , \hspace*{4mm}
\eea
while the annihilation into $ W^3 $ is similar to the abelian one
in eq.~(\ref{sigmaab}) for $e_X =  \cos\theta/2 $.
Note that the cross section is suppressed by the mixing angle as $\cos^4\theta $
and by the fact that the group indices are not summed for the initial state.
Also in this case the rescaled cross section is not just a simple function
of $\beta $, but also of the mass difference in the doublet. We have in fact
\bea
& &  \hspace*{-4mm} \tilde\sigma_{W2} (\beta, \delta^2)  =
2\pi\alpha_2^2 \cos^4\theta\;\beta
\left[
\frac{5}{2} + \frac{11}{6} \beta^2 - \delta^2
+ \frac{4 \beta^2 \delta^4 }{
(1 + 2 \delta^2 )^2- \beta^2}
\right.\nonumber\\
& & \left.
+\; \frac{1 -\beta^2  + 2  \delta^2 -  \delta^4}{\beta}
\; \log \left( \frac{1 + 2  \delta^2  - \beta }{
1 + 2 \delta^2  + \beta} \right)
\right] \, , \hspace*{4mm}
\eea
where $\delta^2 =  (m_{X'}^2 - m_X^2)/s $. The cross section still vanishes for
$\beta = 0$ and is finite for $\delta^2 \rightarrow \infty $.
The detailed expressions for the case of broken EW symmetry are 
much more involved and include also the contribution of the Higgs s-channel allowing for resonance enhancement. They are given in Appendix~B.

\subsection{Sommerfeld enhancement}\label{sect:sommerfeld}

In the previous sections we have computed the annihilation cross sections 
to lowest order in the gauge coupling. 
However, it was shown long ago \cite{Sommerfeld} that an expansion in terms 
of the coupling is inadequate close to threshold, where the velocities of 
the annihilating particles go to zero,
\begin{equation}
\beta \equiv \sqrt{1 -\frac{4 m_X^2}{s}} \rightarrow 0 \,.
\end{equation}
The enhancement at low velocities becomes apparent when one computes
the one-loop corrections, which are enhanced by a factor 
$\frac{C\alpha \pi}{2 \beta}$.
Here, $C$ is a process-dependent constant, $\alpha$ is the gauge coupling
of the annihilating scalars, $\alpha_1$  in the case of $U(1)$ boson
exchanges, or $\alpha_N$ for $SU(N)$ gauge boson exchanges, respectively.
To account for this long-distance effect, one therefore has to resum a
whole class of diagrams, which consist of $t$-channel ladder-type 
exchanges of \emph{massless} soft Coulomb $SU(N)$
or $U(1)$ gauge bosons between the annihilating charged particles.

This resummation of terms $\sim \alpha^n/\beta^n$ leads to the so-called
Sommerfeld factor which multiplies the lowest-order annihilation cross
section. The Sommerfeld enhancement is given by the modulus squared of 
the particle wave function at the origin,
\begin{equation}
E \equiv \left| \Psi(0) \right|^2 = \frac{z}{1-\exp (-z)}, \quad z = \frac{C \alpha \pi}{\beta} \, .
\label{wavefunction0}
\end{equation}
Because this effect is a long-distance one, taking place at a scale 
$\sim \beta m_X$, it factorises from the annihilation cross section 
which is a short-distance effect at the hard-scattering scale 
of order of the mass $m_X$. Schematically,
\begin{equation}
\sigma^{\mbox{\tiny SF}} (\beta, m_X ) = 
E(\alpha(\beta m_X)) \times \sigma^0(\beta) \, .
\label{singlechannel}
\end{equation}
Here, $\sigma^0$ is the leading-order annihilation cross section, 
which has been presented in the preceding subsections.
Eq.~(\ref{singlechannel}) is in principle only valid if the annihilating
partons are in a single $SU(N)$ channel, i.e. for particle in the
fundamental representation either in the singlet or adjoint configurations.
If multiple channels $c$ contribute,
eq.~(\ref{singlechannel}) has to be modified to
\begin{equation}
\sigma^{\mbox{\tiny SF}} (\beta, m_X) 
= \sum\limits_c E_c(\alpha(\beta m_X))
\times \sigma^0_c(\beta) \, .
\label{multiplechannel}
\end{equation}
Here, $\sigma^0_c(\beta) $ is the projection of the leading-order
annihilation cross section in the relevant channel.
For a scalar in the fundamental representation of the $SU(N)$ gauge group
annihilating into massless $SU(N)$ gauge bosons, we have seen that only 
the contribution proportional to the group-symmetric part survives in the
limit of vanishing $\beta $ and is enhanced at low velocities. Therefore 
at leading order the cross sections $ \sigma_c^0 $ can be taken to be the 
same for the singlet and adjoint part up to colour factors and proportional
to the total cross section given in 
eq.~(\ref{eq-sigma-nonabelian})~\footnote{
Taking the true $\sigma_{\mathbf{1}}^0 $ and $\sigma_{\mathbf{A}}^0$
instead, differs from the total $\sigma^0 $ only in the terms suppressed 
by $\beta^2$ and amounts to a correction smaller than $1\% $ at threshold 
where the Sommerfeld factor is effective.}. 
We note also that due to the presence of more than one channel, the 
Sommerfeld factor for an $SU(N)$ gauge theory becomes dependent also 
on the final states, since not all channels may contribute to the 
annihilation into a given final state.

However, the presence of the thermal bath complicates things, as the
interactions with the background gauge bosons may prevent the annihilating
partons to be initially in a definite $SU(N)$ channel.
The time scales for the Sommerfeld effect and the interactions with the 
thermal bath are of competing order, so it is not clear how strong such
effect can be. In this paper we will consider both extreme situations, 
i.e. the case when the thermal bath has no effect and the case when
there is no definite initial channel.
In the latter case, it was argued in the literature that due to the mixing 
of states one should just take an average $ C^{av} $ extracted from the
averaged one-loop correction, leading again to a single Sommerfeld factor 
as in eq.~(\ref{singlechannel}) (see, for example, ref.~\cite{gluinoLSP}). 
While the two approaches give identical results by construction at 
first order, they correspond to two quite distinct resummations of the 
higher orders and they are numerically substantially different.

We obtained the coefficients $C$ 
by computing the $1/\beta$-enhanced contributions for 
$t$-channel $SU(N)$ gauge boson exchange at one loop in the threshold 
expansion (see for example~\cite{threshold} and references therein).
For the generation of the relevant one-loop graphs and the
Lorentz algebra we used the \textit{Mathematica} packages
\texttt{FeynArts} and \texttt{FeynCalc}~\cite{Kublbeck:1992mt}.
We simplified the resulting expressions to only keep terms that are 
leading in $\beta$, that is, we only kept terms that are enhanced in 
the soft region of the one-loop integrals, which were then simple 
enough to perform by hand.
Alternatively, as mentioned above, one obtains the form (\ref{wavefunction0})
directly by computing the normalised wave function at the origin
from the Schr\"odinger equation, describing the annihilating
parton pair, with a Coulomb interaction potential
for positive energies $\sim \beta^2 m_X$~\cite{Sommerfeld}.

The Sommerfeld enhancement due to exchanges of massless
Coulomb $SU(N)$ gauge bosons  is the same for the singlet channel
of annihilation into $SU(N)$ gauge bosons $B_N$ and the annihilation
into $U(1)$ gauge bosons $B_1$,
\beq
C_{S \bar{S} \rightarrow B_N B_N}^{\mathbf{1}} = 
{C_{S \bar{S} \rightarrow B_1 B_1}} = C_F (N) = \frac{N^2-1}{2 N} \, .
\eeq
The factor for the adjoint channel is instead found to be negative and 
thus suppressing,
\beq
   C_{S \bar{S} \rightarrow B_N B_N}^{\mathbf{A}} = 
C_F(N) - \frac{C_A (N)}{2} = - \frac{1}{2N}  \, .
\eeq
The same factors $C^{\mathbf{1}}$ or $C^{\mathbf{A}}$ apply also for 
other final states of the singlet or adjoint channels. 
For example, the Sommerfeld factor for $\tilde t\tilde t^*\to hh$ is 
$C^{\mathbf{1}}_{SU(3)} = 4/3 $, while that for
$\tilde t\tilde t^*\to gh, g\gamma, g Z$ is 
$C^{\mathbf{A}}_{SU(3)}=-1/6$.

Even if the adjoint channel leads to a suppression, upon summing over 
both contributions in eq.~(\ref{multiplechannel}), the net effect is still 
quite enhancing for small $N$.
We have then in fact
\begin{equation}
\sigma^{\mbox{\tiny SF} sum} (\beta, m_X) = \sigma^0(\beta)
\left[ E_{\mathbf{1}} (\alpha(\beta m_X)) \times \frac{2}{N^2 - 2}
+ E_{\mathbf{A}} (\alpha(\beta m_X)) \times \frac{N^2-4}{N^2 - 2}
\right] \, ,
\end{equation}
where, as described above, we have taken 
$ (N^2 - 2)/2 \;\sigma_{\mathbf{1}}^0 = (N^2-2)/(N^2-4)
\; \sigma_{\mathbf{A}}^0 = \sigma^0(\beta) $, and $\sigma^0 (\beta)$ 
is given in eq.~(\ref{eq-sigma-nonabelian}).
For $SU(3)$ this gives
\begin{equation}
\sigma^{\mbox{\tiny SF} sum}_{SU(3)}(\beta,m_X) = 
\sigma^0_{SU(3)}(\beta)  \; \frac{\pi\alpha_3}{42 \beta} \;
\left[\frac{16}{1-e^{-\frac{4}{3}\frac{\pi\alpha_3}{\beta}}}
- \frac{5}{1-e^{\frac{1}{6}\frac{\pi\alpha_3}{\beta}}}
 \right] \, ,
\label{SF-summed-SU3}
\end{equation}
so that the enhancement in the singlet dominates over the suppression
in the adjoint channel.

On the other hand, averaging the one loop contribution over initial 
channels \footnote{Averaging over initial channels is not to be confused
with averaging over initial states which is to be done in addition
when solving the Boltzmann equation.} results in a factor
\beq
{C_{S \bar{S} \rightarrow B_N B_N}} 
= \frac{N^2 +2}{2 N (N^2 - 2)}  =: C_{SU(N)}^{\rm av}\, ,
\eeq
which is although enhancing, much less so than the net effect of the 
summation over singlet and adjoint channels. 
For $SU(3)$, this factor is $C_{SU(3)}^{\rm av} = 11/42$,
leading to 
\begin{equation}
\sigma^{\mbox{\tiny SF} {\rm av}}_{SU(3)}(\beta,m_X) = 
\sigma^0_{SU(3)}(\beta) \;
\frac{\pi\alpha_3}{42 \beta} \;
\frac{11}{1-e^{-\frac{11\pi\alpha_3}{42\beta}}}. 
\label{SF-av-SU3}
\end{equation}
Note that the first term of the expansion of 
eq.~(\ref{SF-av-SU3}) coincides with the 1-loop result 
of \cite{bhsz96} for $gg\to\tilde q\tilde q^*$ near threshold.

\begin{figure}
\centering
\includegraphics[width=14cm]{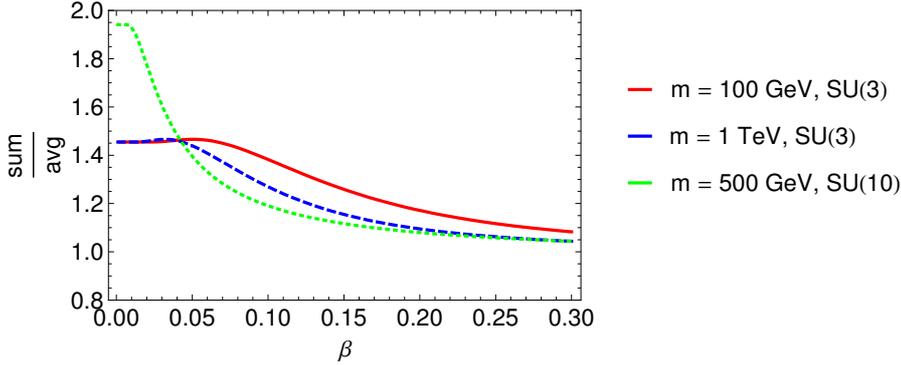}
\caption{\label{fig:sumvsavg}
Ratio of summed over averaged Sommerfeld enhancement, 
$\sigma^{\mbox{\tiny SF} sum}_{SU(3)}/\sigma^{\mbox{\tiny SF}{\rm av}}_{SU(3)}$, 
as a function of $\beta$. The full red line shows the $SU(3)$ case for a mass $m=100$~GeV and 
the dashed blue line for a mass $m=1$~TeV; the dotted green line is for the hypothetical case of 
$SU(10)$ with $m=500$~GeV.}
\end{figure}

If the difference in the exponents in the denominators of 
eqs.~(\ref{SF-summed-SU3}) and (\ref{SF-av-SU3}) could be neglected 
the two expression would be equal.
However, in the small $\beta$ region where the Sommerfeld enhancement 
is relevant, the difference amounts to up to 50 \% for $SU(3)$, and 
is even larger for hypothetical larger $N$, see Figure~\ref{fig:sumvsavg}.

For scalars charged under a $U(1)$ group, there is of course
a corresponding enhancement due to $U(1)$ boson exchanges.
However, the enhancement factor is now governed by the $U(1)$ coupling,
and thus weaker than an enhancement under a strong $SU(N)$ gauge group.
The Sommerfeld factor for the dominant annihilation channel
into $U(1)$ gauge boson pairs can very simply be determined from the 
abelian part of the calculation that led to the factor quoted above. 
We find,
\beq
C_{S \bar{S} \rightarrow B_1 B_1} = 1 \, ,
\eeq
for $t$-channel $U(1)$ exchange,
and the coupling in eq.~(\ref{wavefunction0}) is the $U(1)$ coupling
$\alpha_1$.

Another issue regarding the thermal bath is the fact that gauge bosons 
acquire a mass through interactions with the plasma. 
This Debye screening effect happens at a scale
of order $\sim g T$, whereas the Sommerfeld effect is of order
$\sim \alpha m_X \beta \sim \alpha \sqrt{m_X T} \gg g T$.
Thus the thermal masses of initially massless gauge bosons do not
affect the Sommerfeld enhancement.

Finally, there are also massive gauge bosons such as $W$s and $Z$s 
to consider. The Sommerfeld factor arises from instantaneous
Coulomb exchanges of massless gauge bosons between the slow moving
annihilating pair close to threshold, thus resulting in an $1/\beta$
enhancement, signalling the inadequacy of trying to describe
this exchange in an expansion in terms of loop corrections.
Naturally, massive gauge bosons
have a finite width, and thus cannot be exchanged instantaneously.
In terms of Feynman graphs, the momentum flowing through
a massive gauge boson that is exchanged between the
annihilating pair is naturally cut off by the mass of the exchanged
boson and can never become too soft. The Sommerfeld effect
is exponentially suppressed with the mass of the gauge boson,
as an analysis of the wavefunction picture reveals. 
It can nevertheless become important for relics with masses much 
larger than the electroweak scale, as a very heavy Wino discussed 
in~\cite{Hisano:2006nn}. 
In the following we will consider only the case of massless gauge 
bosons, which is the dominant effect for coloured relics and
for purely right-handed sleptons.
For a more detailed discussion in case of massive EW gauge bosons
we refer the reader to \cite{Hisano:2006nn, Cirelli07}.

\subsection{Unitarity bound}

\begin{figure}[t]\centering
\includegraphics[width=10cm]{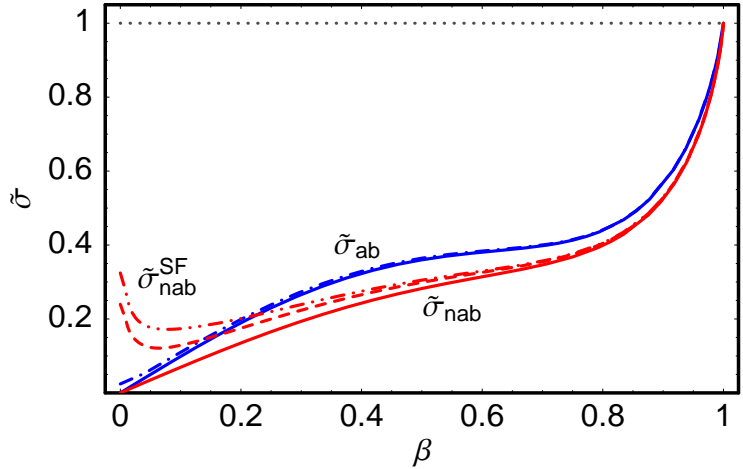}
\caption{\label{fig-sigmas}
Dependence of the rescaled cross sections on the relative velocity
$\beta $, normalised to $1$ at large $s$, i.e. $\beta =1$.
The solid lines show the leading order section, the dashed/dashed
dotted lines the effect of the Sommerfeld enhancement, that makes
the cross sections non-vanishing at the threshold $\beta = 0$.
The $SU(3)$ cross sections are the upper (red) lines, including
the averaged Sommerfeld factor in the dashed line and the
summed one in the dash-dotted. For the abelian case (blue lines)
the Sommerfeld effect is much milder and shown in the dash-dotted line.
Note that the region for $\beta\sim 0$ contributes more strongly
to the thermally averaged cross section due to the Boltzmann-suppression
for large $\beta $.}
\end{figure}

We next compare the above cross sections with the unitarity bound.
Using unitarity and partial wave expansion, the non-elastic cross
section for a particle with spin $s_p$ is given by \cite{unitarity}
\beq 
\sigma_{non-el, J} = 
\frac{4 \pi (2 J+1)(1-\eta_J^2) }{(2s_p+1)^2\;\vec{p}_i^2} 
\eeq 
where $J$ is the angular momentum of
the process, $\vec{p}_i$ is the initial particle momentum, 
$4 \vec{p}_i^2 = s \beta^2 $ in the centre of mass frame in our case,
and $\eta_J^2$ is the contribution of the elastic part. This gives
an upper bound for the annihilation cross section with angular
momentum $J$ as 
\beq 
\sigma_{ann, J} \leq  \frac{16 \pi (2 J+1)}{(2s_p+1)^2 s \beta^2}\; . 
\eeq 
The lowest value is obtained taking $J=0$ and since the s-wave 
annihilation is usually the dominant contribution for a scalar 
non-relativistic particle with $s_p=0$, we will take it as a reference 
value. We therefore have for the maximal rescaled cross section: 
\beq \tilde\sigma_{max} = 16 \pi \label{eq-sigma-unitarity} 
\eeq 
independent of the particle mass or energy. 
In this case the thermal averaging is simple and we obtain 
\beq 
\langle \sigma_{max} v \rangle_x = 
\frac{16\pi}{x^2} \frac{K_2(2x) }{K_2(x)^2} \; , 
\eeq 
which we will consider in the following to be the maximal cross section 
per degree of freedom\footnote{ Note that here we are computing
explicitly in the centre of mass frame, while the Boltzmann
equation requires to use the covariant or lab frame. 
The difference between the two frames has been discussed in
\cite{gg91} and gives only a small correction for non-relativistic
particles, which we neglect here.}.
We see clearly that the cross sections discussed above
satisfy this bound and are suppressed at the very least
by $\alpha^2$.
Figure~\ref{fig-sigmas} shows the rescaled cross sections
for the abelian and non-abelian cases,
eqs.~(\ref{eq-sigma-abelian}) and (\ref{eq-sigma-nonabelian}),
together with the unitarity  bound eq.~(\ref{eq-sigma-unitarity})
as a function of the relative velocity of the annihilating particles.

The unitarity cross section $\tilde\sigma_{max}$ can be used to obtain
a lower bound of the yield. Moreover, it can be taken as the maximal
annihilation cross section possible even after the QCD phase transition,
when the coloured
states are confined into the equivalent of scalar hadrons and fermionic
mesons~\cite{QCDconfined}.
Constraints from cosmology on such kind of hadronic states have been mostly
studied for the case of a stable exotic quark~\cite{Nardi:1990ku}, a
gluino LSP~\cite{gluinoLSP} or for very long-lived
gluino in the split SUSY scenarios~\cite{gluino-splitsusy}.
It has been argued in \cite{Kang:2006yd} that the annihilation cross section
for such states could become much stronger, if bound states between two scalar
hadrons/fermionic mesons are formed with rate $ \sim \pi/\Lambda_{QCD}^2 $
and in that case the coloured relic abundance after the QCD phase transition 
is further reduced below $ Y \sim 10^{-16}-10^{-17} $.
We will not consider this possibility in the following, but note however 
that, while most of the cosmological bounds for a decaying relic are then 
satisfied, one still needs to consider the bounds for
a stable relic.

\subsection{Thermally averaged cross sections and velocity expansion}

We integrate eq.~(\ref{sigmatermica}) numerically  to
obtain the thermally averaged cross section.
Very often such a quantity is instead approximated with
the first terms of its velocity expansion, since the
relevant regime takes place when the annihilating particles
are already non-relativistic.
To obtain such an expansion, one can use the approximation
\begin{equation}
s - 4 m_X^2 \simeq 4 m_X^2 \beta^2
\end{equation}
and expand in $\beta $ the expression
\begin{equation}
\sigma v_{\rm M\o l} \simeq {1\over 2 m_X^2 \beta }
\tilde\sigma \left(\beta \right) \; .
\end{equation}
We see that if $\tilde \sigma $ is constant at zero velocity,
the cross section is enhanced like $1/\beta $ in that limit.
This is indeed the case both for the Sommerfeld-enhanced
cross section and the unitarity one.

The first term in the expansion, which is independent of the
velocity and coincides therefore with the first term in the expansion
of the thermally averaged cross section \cite{gg91},
is given by
\begin{eqnarray}
\sigma_{ab} v
 & \rightarrow  &{2\pi\alpha_1^2 e_X^4 \over m_X^2} + {\cal O}(\beta^2)\;,
\\
\sigma_{nab} v
& \rightarrow &
\frac{\pi\alpha_N^2 }{m_X^2}
\frac{(N^2-1)(N^2-2)}{4 N^3} + {\cal O}(\beta^2)\; ,
\end{eqnarray}
for the abelian and non-abelian cases respectively.

\begin{figure}[t]\centering
\includegraphics[width=10cm]{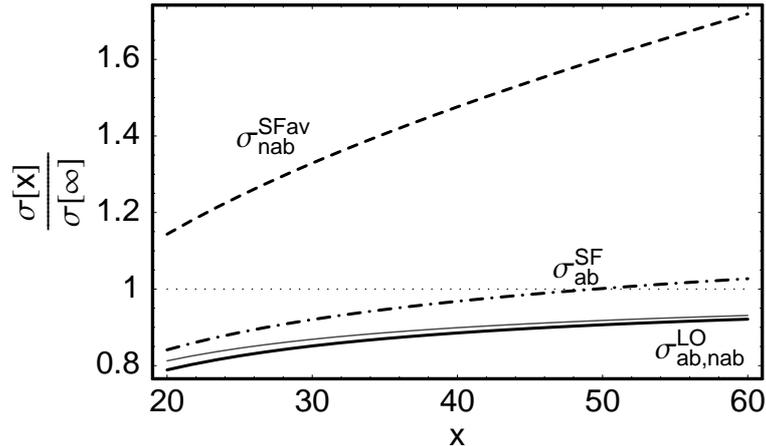}
\caption{\label{sigmath}
Ratio of the thermally-averaged cross section and the first term
in the velocity expansion around $\beta = 0$, for $m_X = 350$ GeV.
The thick solid line is for the abelian, the thin line for the
non-abelian ($SU(3)$) case. Dash-dotted and dashed respectively are
the same ratios including the Sommerfeld enhancement, only the
averaged one for the non-abelian case: we see that in this case the
thermally averaged cross sections do not converge to the first order
term in velocity, but that the latter can still give a good estimate
within 15\% of the full result in the abelian case;
for the non-abelian case the Sommerfeld enhancement changes the result
considerably and the velocity expansion fails. Note that the case
of the summed Sommerfeld factor is outside the range of the plot.
}
\end{figure}

We plot in Figure~\ref{sigmath} the thermally averaged cross sections as
a function of $x$ normalised with respect to the first term in their
velocity expansion including
also the Sommerfeld enhancement factor, both for the abelian case and for the
QCD case with $N=3$.
We see that keeping only the lowest order overestimates the thermally averaged
cross section, i.e. underestimates the yield,  in the abelian case
by at most $20\%$ in the region of freeze-out ($x\sim30$).
The non-abelian case for $N=3 $ is approximated better also because
the freeze-out takes place at a larger $x\sim40$, i.e. smaller $\beta $.
On the other hand, once we include the Sommerfeld enhancement, the
thermally averaged cross section does no more converge to the 
first constant term in the velocity expansion due to the threshold 
singularity at $\beta = 0 $.
Nevertheless the first order term {\it without } the enhancement can still
give a reasonably good approximation for the abelian case, since the
Sommerfeld enhancement partially compensate the $ 20\%  $ underestimation
of the Born result.
For the non-abelian case the Sommerfeld enhancement is so strong that
the low energy expansion can give only an order of magnitude estimate.

\section{Results for the relic density}
\label{results}

We solve the Boltzmann equation (\ref{Boltzmann-2}) numerically
for the exact thermally averaged cross sections given above.
This improves the old results \cite{wolfram} that were obtained
with the velocity expansion.

For the case of an abelian charged relic, we consider $e_X = \pm1$
and we set the coupling to be $\alpha_{em}=1/128$.
For the non-abelian case we take $N=3$ and $\alpha_N$ to be the QCD 
coupling $\alpha_3(Q)$ with $Q=2m_X$ in the hard process and 
$Q=\beta m_X$ in the Sommerfeld correction,
c.f.\ Sect.~\ref{sect:sommerfeld}. In order to avoid the non-perturbative 
regime, we cut off the running of $\alpha_3$ at $Q=2$~GeV, 
i.e.\ $\alpha_3(Q<2~{\rm GeV})\equiv \alpha_3(2~{\rm GeV})$.

For the entropy and energy density parameters we take
$g_S^{1/2}=g_{\rho}^{1/2} = 10$, since we expect the freeze-out
to take place between 10--100 GeV, when only the light
Standard Model particles are still in equilibrium in the thermal bath.

\begin{figure}\centering
\includegraphics[width=10cm]{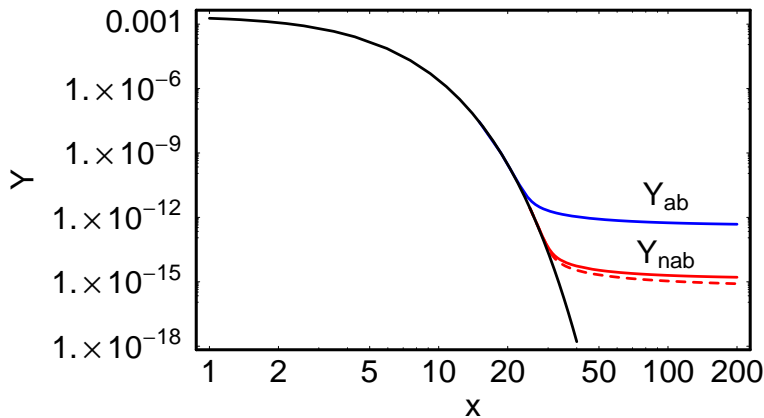}
\caption{\label{freeze-out}
Time evolution of the particle yield for the cases of abelian
and non-abelian cross section, for $m_X = 200$ GeV. 
The upper (blue) curve is for
an electromagnetically charged scalar particle with unit charge,
while the lower (red) curves correspond to a single coloured scalar in the
fundamental representation without the Sommerfeld factor (solid) and
with the Sommerfeld factor averaged (dashed).
We see that the treatment of the Sommerfeld factor has an impact
of about 30\% on the final number density. }
\end{figure}

Our results are plotted in Figure~\ref{freeze-out}.
We see that the yield $Y$ follows relatively closely the equilibrium
density until the time of freeze-out, which happens at different
values of $x$ for the different cross sections.
As expected the non-abelian interactions being stronger gives a
considerably lower relic density. The ratio between the two cases is well
approximated by the ratio of cross sections, 
$\sigma_{nab}v/\sigma_{ab}v$,
at zero velocity :
\beq
{Y_{ab}\over Y_{nab} }
= {7\over 27}\; {\alpha_{3}^2 \over \alpha_{em}^2} \approx 40\,.
\label{ratioY}
\eeq

We next consider the dependence on the only dimensional parameter,
the mass of the charged relic. We have seen that the thermal
average can be written only as a function of $x$ and since we are
integrating the Boltzmann equation to $x \rightarrow \infty $ we
get rid of the dependence on $m_X$ that is contained there. A
subleading dependence would survive by integrating to a finite
value of $x$, but this effect is negligible for the present universe
with a temperature $ T_{now} \sim 10^{-4} \mbox{eV} \ll m_X $. On
the other hand, the mass directly enters in the coefficient of
eq.~(\ref{Boltzmann-2}) and that is the stronger dependence on
$m_X$.  Note that this dependence is present even in the unitarity
case, where the reduced cross section is explicitly independent
of the mass and velocity. In general therefore the yield is
proportional to the mass and can be rescaled as 
\beq 
Y(m_X) = Y(1\,\mbox{TeV}) \left({m_X \over 1\,\mbox{TeV}} \right)\; . 
\eeq
with $ Y_{ab} (1\,\mbox{TeV}) = 3.9\times10^{-12}$ and $ Y_{nab}
(1\,\mbox{TeV}) = 1.6\times10^{-13}$ for the abelian and non-abelian
cases, respectively, for the total degrees of freedom, including
antiparticles.
For the case of the unitarity cross section, the total yield becomes 
instead $ Y_{lim} (1\,\mbox{TeV}) = 6.6 \times 10^{-18} $ 
(or $ 2 \times 10^{-17} $ for three degrees of freedom).

\begin{figure}\centering
\includegraphics[width=10cm]{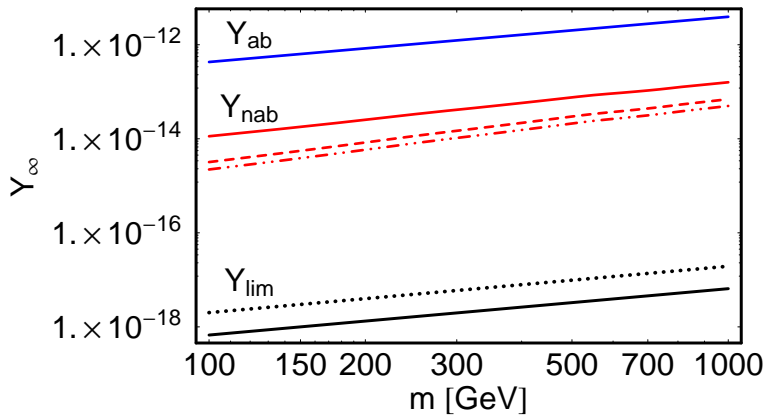}
\caption{\label{Yofm}
Dependence of the yield on the mass of the charged relic. 
From top to bottom, the first (blue) line is for the case of an 
electromagnetically charged relic, while the second (red) line is 
for a coloured relic, the dashed and dash-dotted lines include 
the Sommerfeld factor, averaged and summed respectively.
The lower two (black) lines correspond to the maximal annihilation 
cross section given by unitarity --
the solid one for  a single d.o.f., the dotted one for $3$ d.o.f.\
for the fundamental representation of QCD.
Note that the non-abelian case is still three orders of magnitudes
away from the unitarity cross section.}
\end{figure}

Since the energy density also increases for larger masses, this can 
be used to give a constraint on the mass of any stable thermal relic 
from the maximal cross section allowed by unitarity \cite{unitarity}.
Using the WMAP 5-year results~\cite{WMAP5} for the most conservative 
upper bound for the matter density, we can update such bound. 
In fact imposing
\begin{equation}
\Omega_X h^2 = m_X Y_{X+\bar{X}} (T_{now}) s(T_{now})/\rho_c
\leq 0.13
\end{equation}
gives us for a single degree of freedom the constraint
\begin{equation}
\frac {m_X Y_{X+\bar{X}} (T_{now}) }{ \mbox{GeV}}
\leq 4.6 \times 10^{-10}
\end{equation}
resulting for a scalar particle in
\begin{equation}
m_X \leq 280\; \mbox{TeV}\; .
\end{equation}
Note that for a fermionic spin 1/2 relic the unitarity cross section
is reduced by a factor four and therefore the bound on the mass is 
stronger by a factor two.

\section{Constraints on cosmological relics}

We review here the constraints on the abundance of cosmological relics that
we will compare with the number density of a charged scalar relic in the
next section.
First we will consider the case of stable relics (i.e. with lifetimes longer
than $10^{27} $ s) and next relics with lifetimes in the window
$ 0.1 - 10^{12} \mbox{s} $.
Note that for shorter lifetimes the constraints are non-existent, as long
as the particle did not dominate the universe dynamics before decaying or
produce a large amount of entropy, while for lifetimes between
$10^{10}-10^{27} \mbox{s} $ bounds from CMB distortion \cite{Lamon:2005jc} and from the measured
photon diffuse flux \cite{photon-flux} apply, but will not be discussed here.

\subsection{Stable relics}

The possibility of existence of some more exotic cosmological relics than the
known light elements stimulated many years ago the search for exotic nuclei
in water and other materials on the earth. Those searches were unsuccessful
and provide a very strong limit on the number density of any relic that would
bind electromagnetically with an electron or in nuclei, under the assumption
that such particles are equally distributed in the Universe compared to baryons.
If such relics were present long before structure formation, it is highly probable
that they were trapped together with baryons when the universe's density was
still nearly homogeneous, so that we can expect their number density
not to be too strongly dependent on the local environment. Note that in
any case these bounds are so strong that the possibility of such a relic to
be Dark Matter is completely excluded.

The most recent constraints are those obtained by \cite{Yamagata:1993jq} looking
for anomalously heavy hydrogen in deep sea water, which apply to an electrically
positively charged relic, and give for masses  $ 5\; \mbox{GeV} \leq m_X \leq 1.6\; \mbox{TeV} $:
\beq
Y_{X^+} \leq 4 \times 10^{-17}\, Y_B =
3.5 \times 10^{-27} \left( \frac{ \Omega_B h^2 }{0.0223} \right)
\eeq
Taking into account the gravitational effect in deep sea,  this corresponds to
a concentration of the order of $ 10^{-28} $ at sea level or equivalently
\beq
Y_{X^+} \leq 0.9 \times 10^{-38} \left( \frac{ \Omega_B h^2 }{0.0223} \right) \; ,
\eeq
which is comparable to other limits in the same mass range, \cite{Smith:1982qu}.
For larger masses up to a TeV,  a slightly looser bound
$ Y_{X^+}/Y_B < 3 \times 10^{-20} $ was found by \cite{Hemmick:1989ns},
while for even larger masses $ 10\; \mbox{TeV} \leq m_X \leq 6 \times 10^4\; \mbox{TeV} $
it weakens even further to $ Y_{X^+}/Y_B < 7 \times 10^{-15} $, as given by
\cite{Verkerk:1991jf}, i.e.
\beq
Y_{X^+} \leq 6 \times 10^{-25}  \left( \frac{ \Omega_B h^2 }{0.0223} \right) \; .
\eeq

For electromagnetically neutral, but coloured relics, the bounds are obtained
from considering heavier elements and are considerably weaker; using the
results of \cite{Hemmick:1989ns} for Carbon, the limits are of the order
$ Y_{X^+}/Y_B \leq 4-8 \times 10^{-20} $ for $ m_X = 0.1-1\; \mbox{TeV} $, reaching
$ 2 \times 10^{-16} $ at the largest mass considered $ 10\; \mbox{TeV} $.
For larger masses $ m_X \leq 100\; \mbox{TeV} $ only the constraint by
\cite{Norman:1988fd} for lead is present, giving
\beq
Y_{X} \leq 1.5 \times 10^{-13}\, Y_B =
1.3 \times 10^{-25} \left( \frac{ \Omega_B h^2 }{0.0223} \right) \; .
\eeq

We see that these constraints are very strong. In order to reach even the weakest
bound of $ Y_X \leq 10^{-25} $,  the unitarity cross section is way too weak
and needs to be increased at least by nine orders of magnitude, i.e.
\beq
\sum_J (2 J + 1) > 10^9\; .
\eeq
Therefore stable relics are allowed only if their interaction does not belong to the
Standard Model and they cannot form exotic atoms/nuclei or if their annihilation
rate becomes much larger than the unitarity one as it can happen if they interact
strongly and can form intermediate bound states. But in any case, note that
cross sections of the order $ \pi/\Lambda_{QCD}^2 $ that can arise after the QCD
phase transition are not sufficient to evade these constraints \cite{Kang:2006yd},
so their interaction would have to be stronger than QCD.

\subsection{Unstable relics}

Different cosmological constraints exist on the density of an unstable relic,
depending on its lifetime.
For lifetimes between $0.1$\,s and $10^{10}$\,s,  the
strongest constraints come from Big Bang Nucleosynthesis.
In fact, if the relic decay injects very energetic particles into the thermal
bath during BBN, it can change the abundances of the light elements.
Since standard BBN agrees quite well with the primordial abundances of
Helium-4, Deuterium and (within a factor of two) Lithium-7 inferred from
present astronomical observations~\cite{BBNrev},  the relic density has to
be low enough not to change those predictions too strongly.
These effects are present for any decaying particle and have been studied
in various papers (see \cite{neutBBN,BBN-kkm,Jedamzik:2006xz,Kawasaki:2008qe} and 
references therein).
For lifetimes above $ 3000 \;\mbox{s} $, corresponding to the time of
production of Lithium, additional constraints are present if the relic is
electromagnetically charged and can form a bound state with positively
charged nuclei increasing the rates for Lithium-6 production~\cite{CBBN}.
The Standard BBN prediction for the $^6Li $ abundance is actually way too
small compared to the observed one, so that the presence of a charged relic
with appropriate lifetime can help reconciling BBN with the measured
abundances of $^6 Li,\, ^7Li $~\cite{Li6}, but we will disregard this
possibility and only concentrate on the exclusion region.

We summarise here the main results from various BBN analyses and give
conservative bounds on the energy density of the decaying relic and compare
them with our computation of the relic density.
Since we are interested in escaping the BBN constraints, we focus mainly
on the strongest bounds, but we keep conservative values for the light
element abundances.
Note that in many of the analysis slightly different ranges for these
abundances are considered, corresponding to slightly different constraints
on the decaying relic.

In general, the decay can produce very energetic SM particles that can
initiate either hadronic or electromagnetic showers in the plasma.
The most stringent bounds are obtained for a relic that produces mostly
hadronic showers, since electromagnetic particles like photons or electrons
can thermalise very quickly by interacting with the tail of the 
CMB distribution until times of about $10^6$ s.
So we will consider in the following the constraints for relics producing
hadronic showers with a branching ratio $B_H = 1$.
We will comment later on the case where this branching ratio is smaller.
There are then practically three regions of the lifetimes as discussed in
\cite{BBN-kkm}:

\begin{itemize}
\item{$10^{-1}\,\mbox{s} \leq \tau \leq 10^2 \,\mbox{s} $ : the dominant
effect is the interconversion between protons and neutrons,
that changes the Helium abundance, overproducing it;
}
\item{$10^2\,\mbox{s}  \leq \tau \leq 10^7 \,\mbox{s} $: hadrodissociation
is the most efficient process and the bound come from the
non-thermal production of $Li $ and $D$;}

\item{$10^7\,\mbox{s}  \leq \tau \leq 10^{12} \,\mbox{s} $: photodissociation
caused both by direct electromagnetic showers and by those generated
by the daughter hadrons starts to dominate and the overproduction
of $^3He$ is the main result.
}
\end{itemize}
It is clear that these limits depend on the decay branching ratio $B_H$
into hadrons for lifetimes $\tau \leq 10^7 \;\mbox{s} $, while they are
independent of $B_H$ for longer lifetimes.
In Table~\ref{tableBBN}, we give conservative bounds taken from
the general analysis  of \cite{BBN-kkm} for the three regions,
assuming $B_H=1$. Similar
constraints were obtained independently also by \cite{Jedamzik:2006xz}. Note that the bound for short lifetimes becomes
approximately one order of magnitude weaker if one takes a more recent
value of the $^4 He$ abundance as discussed in \cite{Kawasaki:2008qe}. Unfortunately this new publication does
not provide constraints for a general relic, but discusses only the
explicit cases of a bino neutralino or a right-handed stau.

The limits we use can be parameterised as
\begin{eqnarray}
  Y_{X+\bar{X}} \leq 1.0 \times 10^{-13}\, 
          \left({m_X\over 1 \mbox{TeV}} \right)^{-0.3}
          && \mbox{for}\;\; \tau_X \sim 0.1-10^2\, \mbox{s}\,, \\
  Y_{X+\bar{X}} \leq 1.1 \times 10^{-16}\, 
          \left({m_X\over 1 \mbox{TeV}} \right)^{-0.57}
         && \mbox{for}\;\; \tau_X \sim 10^2-10^7\, \mbox{s}\,.
\end{eqnarray}
The assumption $B_H=1$ is surely valid if the decaying relic is coloured,
while $B_H$ can be  different if it is only electromagnetically charged,
as in the case of the stau.
If the branching ratio into hadronic modes for the relic is less than one,
the hadronic BBN bounds are relaxed accordingly by a factor $1/B_H$.
For intermediate lifetimes, then electromagnetic showers can become a
more important effect, but only if $B_H < 0.01 $.

\begin{table}[t]
\begin{center}
Maximal values of $ m_X Y_{X+\bar{X}} $ (GeV) allowed by BBN\\[2mm]
\begin{tabular}{| c | c | c | c |}\hline
  & & & \\
  $m_X$ (TeV) &
  $10^{-1}-10^2 $ s &  $10^2-10^7 $ s & $10^7-10^{12} $ s \\
  & & & \\
\hline \hline
& & & \\
$0.1$ & $ 2 \times 10^{-11} $  & $ 5\times 10^{-14}  $ & $ 10^{-14} $  \\
 & & & \\
\hline
& & & \\
 $1$ & $ 1 \times 10^{-10} $  & $ 10^{-13} $  & $ 10^{-14} $ \\
& & & \\
\hline
& & & \\
 $10$ & $ 5 \times 10^{-10}  $  &  $ 3 \times 10^{-13} $ & $ 10^{-14}$ \\
& & & \\
\hline\hline
\end{tabular}\\
\end{center}
\caption{Maximal allowed values of $m_X Y_{X+\bar{X}}$ in the 
different
region of lifetimes taken from Figures~38--40 of \cite{BBN-kkm}.
We are assuming here that the energy released in Standard Model
particles is one half of $m_X$ as happens in a two body decay of
the NLSP into LSP and the NLSP non-supersymmetric partner and
that all the energy is released in hadrons.
In general the strongest bound is for longer lifetimes and
it is independent of $m_X$ and the hadronic branching ratio.
The bounds in the second column come from $D$, but the
$^6Li$ ones, that are sometimes considered too strong~\cite{Li6},
are not very far away.
\label{tableBBN}}
\end{table}

For electromagnetically charged relics with lifetimes longer than about
$3000\;\mbox{s}$ and low $B_H < 0.1-0.01 $,  strong bounds also come from
considering the catalysed overproduction of $^6 Li$ \cite{CBBN}.
In fact when bound states between nuclei and the relic can form such as
$ ^4He X^{-} $, many nuclear rate are modified and change the final
abundance especially of $^6Li$ and $^7 Li$.
For particles decaying after  $5 \times 10^5 \mbox{s} $ it has been argued that
uncertainties in the nuclear rates make such constraints weaker than the 
general ones discussed above~\cite{Jedamzik:2007qk}, so we will consider here
catalysed BBN constraints only for the intermediate lifetime range.

Unfortunately, different values for these bounds are given in the literature;
in \cite{Hamaguchi:2007mp, Pradler:2007is} they are found to be maximally
at the level of $ Y_{X^-} < 1.4$\,--\,$2  \times 10^{-16} $, while the 
latest value in \cite{Jedamzik:2007qk} is maximally $ Y_{X^-} < 10^{-14} $, 
taking a larger window for the ratio $^6Li/^7Li $.
Here we will use as a constraint the simple interpolation for the total 
yield~\footnote{The catalysed BBN
 constraints restrict only the abundance of the negatively charged particles,
 but we give here the constraint for the total yield assuming
 $ 2 Y_{X^-} = Y_{X+\bar{X}} $.}
 \begin{equation}
 Y_{X+\bar{X}} \leq  \left\{
 \begin{array}{ll}
 2 \times 10^{-12} \left( \frac{\tau_X }{3 \times 10^3 \mbox{s} }\right)^{-2}
 & \mbox{for}\quad\quad \tau_X \lsim 10^5 \mbox{s} \cr
 2 \times 10^{-15}   & \mbox{for}\quad\quad \tau_X \geq 10^5 \mbox{s}
 \end{array}
 \right.
 \end{equation}
 that lies somewhat in between.
The bounds from catalysed BBN do not apply for coloured scalar relics because 
these should have a large branching ratio into hadrons, such that the 
`conventional' BBN bounds from hadronic showers are much stronger.  
In passing note also that  up-type squarks would mostly hadronise 
into neutral fermionic mesons which are lighter than the charged 
ones~\cite{QCDconfined}.

\begin{figure}\centering
\includegraphics[width=8cm]{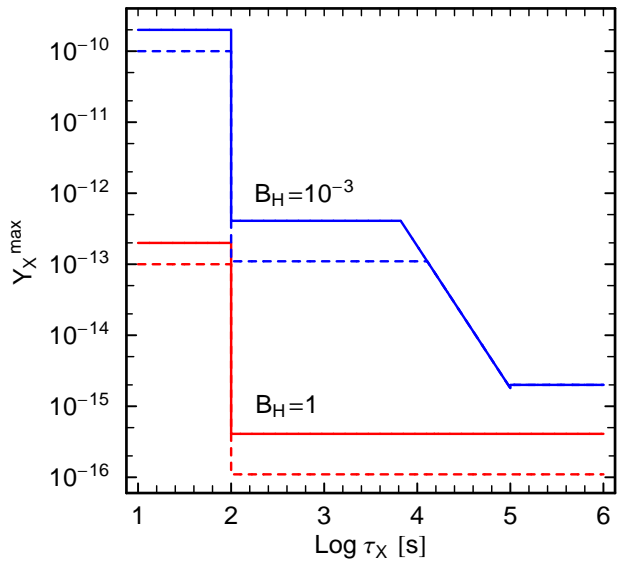}
\caption{\label{BBNlimits}
Maximal total yield $Y_{X+\bar{X}}^{\rm max}$ allowed by BBN as a 
function of the relic's lifetime $\tau_X$ for the two cases 
$B_H=1$ (red) and $B_H=10^{-3}$ (blue). 
The full lines are for a mass of $m_X=100$~GeV, while the dashed lines 
are for $m_X=1$~TeV. Note that for $B_H=10^{-3}$,
the limit for $\tau_X\gsim 10^4$\,s comes from CBBN.}
\end{figure}

We summarise the constraints in Fig.~\ref{BBNlimits}, which shows our
conservative bounds in the plane of total number density vs lifetime.
Note that the constraint from catalysed BBN are for the stau stronger 
than the hadronic ones for lifetimes longer than $ \sim 10^4 $\,s and 
exclude a light stau NLSP with a 100 GeV gravitino LSP in the 
CMSSM~\cite{stauNLSP}.

Comparing with Fig.~\ref{Yofm}, we see that even for a charged relic 
that can annihilate efficiently, the BBN bounds are very strong;
in particular the case of a simple abelian interaction seems to be
excluded for any charged relic whose lifetime is longer than 0.1 s
and produces hadronic showers with $B_H=1$.
For the coloured case the situation is less severe, but even 
with the Sommerfeld enhancement, which reduces the yield substantially, 
it is not possible to evade the bounds completely.
Still all masses above approximately $50 $ GeV are excluded for
lifetimes longer than 100 s, while for shorter lifetimes masses
up to 700 GeV are allowed.
A much larger number of colours than three would be needed to relax 
all bounds.  
Even the unitarity case reaches the strongest BBN constraint at masses
around 700 GeV for $3$ degrees of freedom or 1 TeV for a single one.

\section{Application to the MSSM}

Until now we have considered the ideal case that the relic particle has 
only one single interaction. In realistic models, however, more than 
one interaction -- and hence more than one annihilation channel -- 
is present, making the BBN bounds less stringent.

In this section, we discuss the concrete examples of a relic stau or 
stop in the MSSM. We use the MICROMEGAS package \cite{micro} to take 
into account all relevant annihilation and co-annihilation channels, but compare also with 
the results for $Y_{ab}$ or $Y_{nab}$ for the case of one single gauge 
interaction.

\subsection{Relic stau}

Our results for an electrically charged relic can be applied, for instance,
to the case of the supersymmetric partner of the $\tau$.
We assume here that the relic stau is a right-chiral state, $\tilde\tau_R$,
and that all other SUSY particles as well as the heavy Higgs bosons decouple.

\begin{figure}[t]\centering
\includegraphics[height=10cm]{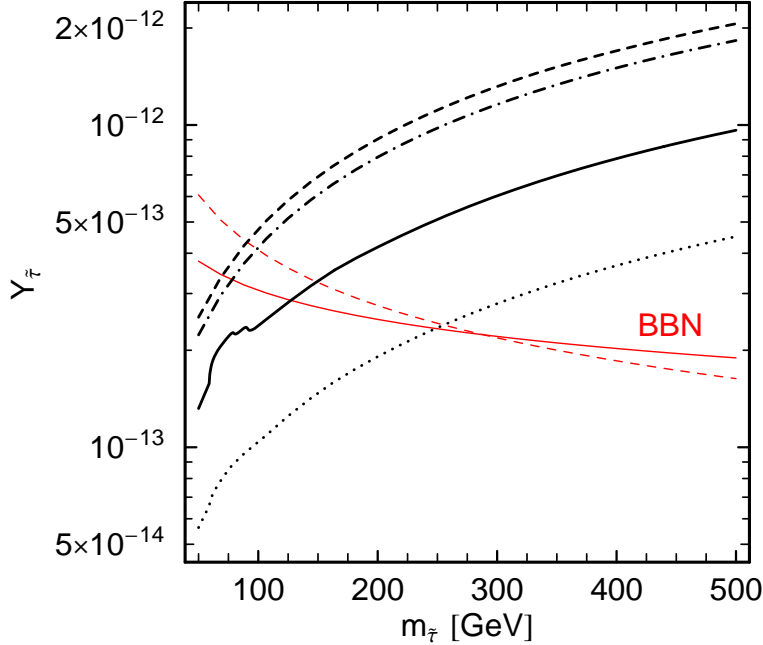}\vspace*{-4mm}
\caption{Total yield $Y$ (black lines) for a relic $\tilde\tau_R$ as
function of the stau mass.
The dashed and dashdotted curves are tree-level and Sommerfeld-
corrected results, respectively, from
annihilation into photons (i.e.\ $Y_{ab}$). The full line includes
also annihilation into
$W$ and $Z$ bosons, assuming all other sparticles decouple.
Finally, the dotted line shows the case $m_{\tilde B}=1.1m_{\tilde
\tau_1}$.
BBN bounds are shown in red: as full line for 0.1--100\,s lifetime and
$B_H=0.65$,
and as dashed line for $>\,100$\,s lifetime and $B_H=10^{-3}$.
Note that if the lifetime exceeds about $10^4$\,s, the CBBN
constraints become more important and quickly exclude number densities
at the level of $10^{-13}$--$10^{-15}$, see Fig.~\ref{BBNlimits}.
\label{YstauNum}}
\end{figure}

The dependence of the yield on the stau mass is shown in Fig.~\ref{YstauNum}.
For a 100 GeV $\tilde\tau_R$, we get $Y_{\tilde\tau}=4.8\times 10^{-13}$ at 
tree level from annihilation into photons (c.f.\ the dashed line). 
This is reduced by about 12\% by the Sommerfeld enhancement (dashdotted line). 
In the full EW theory, the stau also annihilates into $W^+W^-$, $ZZ$ and 
$\gamma Z$.
In fact, for $m_{\tilde\tau}=100$~GeV, the $\gamma\gamma$ channel contributes
about 55\%, $\gamma Z$ about 25\%,
$ZZ$ about 10\% and $WW$ about 5\% to the total rate; the remaining 5\%
go into SM fermions. At higher stau masses, we have $\sim$50\% $\gamma\gamma$
and $\sim$30\% $\gamma Z$.
Overall this gives a reduction of $Y$ by a factor of about 2 (solid line), 
leading to $Y_{\tilde\tau}=2.4\times 10^{-13}$ at $m_{\tilde\tau}=100$~GeV.

Staus can also annihilate into $\tau\tau$ through $t$-channel neutralino 
exchange. We here consider only the bino contribution. Lowering the bino 
mass $m_{\tilde B}$ decreases the yield until bino-stau coannihilation takes 
over, increasing it again. We find a minimum yield at about 
$m_{\tilde B}\simeq (1.1-1.2)m_{\tilde\tau_1}$, shown as dotted line in Fig.~\ref{YstauNum}. 
It is roughly a factor 2 lower than the solid line, in agreement with \cite{ahs00}. 
Note also that the neutralino exchange leads to annihilation of same-sign stau pairs, 
$\tilde\tau_1^\pm\tilde\tau_1^\pm \to \tau^\pm\tau^\pm$, so this process gets Sommerfeld-{\it suppressed},   
and the total Sommerfeld effect almost cancels.

The annihilation into $W^+W^-$ and $ZZ$ is considerably enhanced if the
relic stau also has some $\tilde\tau_L$ component,
$\tilde\tau_1=\tilde\tau_R\sin\theta + \tilde\tau_L\cos\theta$ with 
$\cos\theta\not=0$.
In this case also t-channel exchange of $\tilde\nu_\tau$ (for $W^+W^-$)
and $\tilde\tau_2$ (for $ZZ$) has to be taken into account in addition 
to the 4-vertex
and $s$-channel $\gamma/Z$ exchange, c.f.\ Appendix B. It turns out that these
$t$-channel diagrams lead to a destructive interference: for given 
$\cos\theta$, smaller $\tilde\nu_\tau$ and $\tilde\tau_2$ masses lead 
to smaller cross sections.
Since the stau and sneutrino masses and stau mixing angle are related to
each other, one cannot simply maximise the cross section by choosing 
maximal mixing ($\cos\theta=0.7$) and very heavy $\tilde\nu_\tau$ and 
$\stau_2$. However, for reasonable parameter choices, it is still 
possible to reduce the yield shown in Fig.~\ref{YstauNum} by up to about 
an order of magnitude.
Alternatively, one could rely on resonant annihilation through 
\linebreak
s-channel Higgs exchange or on coannihilation with sparticles
that are close in mass to bring $Y_{\tilde\tau}$ below the BBN bounds.

Barring these possibilities of largely enhanced cross sections, the stau 
lifetime and branching ratio into hadronic modes become key parameters 
to decide whether the scenario is allowed.
First of all, let us discuss briefly the branching ratio into
hadrons. We are considering here the decay
$\tilde\tau_R\to\tau$+LSP. The $\tau$ decays into charged mesons
65\% of the time, while the remaining times into leptons only.
Charged mesons have a similar effect as nucleons during BBN only
at short times $ < 100$ seconds, because later they decay before
interacting with nucleons and give rise only to electromagnetic
showers \cite{BBN-kkm}. Therefore we will take
$B_H(\tilde\tau)\sim 0.65$ for lifetimes up to 100 s, while it
becomes much smaller for longer lifetimes, we will use 
$B_H(\tilde\tau)\sim 10^{-3}$ as reference value.
This is in the central range computed recently for the stau decay
into tau, gravitino and a $q\bar q$ pair, and we refer to
that result for a more detailed analysis~\cite{frank06}.
(A full computation including a more complete treatment of the 
hadronic decays of the tau for the case of a right-handed
stau has been given in~\cite{Kawasaki:2008qe}.)
We have then to apply the BBN bounds discussed in the previous
section corrected by these branching ratio factors, according
to the time of decay.

Regarding the stau lifetime, this depends strongly on the nature of the LSP.
For the case of the axino LSP, the decay rate is given by
\beq
\Gamma (\tilde\tau_R\rightarrow \tau \tilde a)
= (25\;\mbox{s} )^{-1} \xi^2
\left({m_{\tilde\tau}\over 10^2\;\mbox{GeV}} \right)
\left({m_{\tilde B}\over 10^2\;\mbox{GeV}} \right)^2
\left( {10^{11}\;\mbox{GeV}\over f_a} \right)^2
\left(1 - {m_{\tilde a}^2\over m_{\tilde\tau}^2} \right)
\eeq
where $m_{\tilde a}$ is the axino mass, $m_{\tilde B}$ is the Bino mass,
$f_a$ is the Peccei-Quinn scale, and
$\xi $ is a factor of order 1 taking into account
some uncertainties in the loop computation \cite{bchrs05}.
Therefore only the weakest BBN bound applies and actually
disappears completely for large stau mass: in fact even
for the conservative case $ m_{\tilde B} = 1.1\,m_{\tilde \tau} $
and $ f_a = 10^{11} $ GeV, the lifetime becomes shorter than 0.1\,s for
$m_{\tilde\tau} \leq 590$~GeV. We are here neglecting the case of a 
strong degeneracy between the stau and axino masses.
We see therefore that for axino LSP a very light stau is a viable
possibility and, depending on the supersymmetric spectrum, only 
the mass window between $125/250-590 $ GeV is possibly excluded by
the BBN constraints, as can be seen from Fig.~\ref{YstauNum}. 
In that region however probably a more proper computation of the 
stau hadronic branching ratio and its effect in the early stages
of BBN is needed, as discussed in \cite{Kawasaki:2008qe}.
In fact comparing our exclusion region with theirs, we find that 
their constraints are much weaker for short lifetimes, due to an 
up-dated value of the Helium abundance and a larger systematic error,
allowing all the stau region for an axino LSP.

\begin{figure}[t]\centering
\includegraphics[height=7cm]{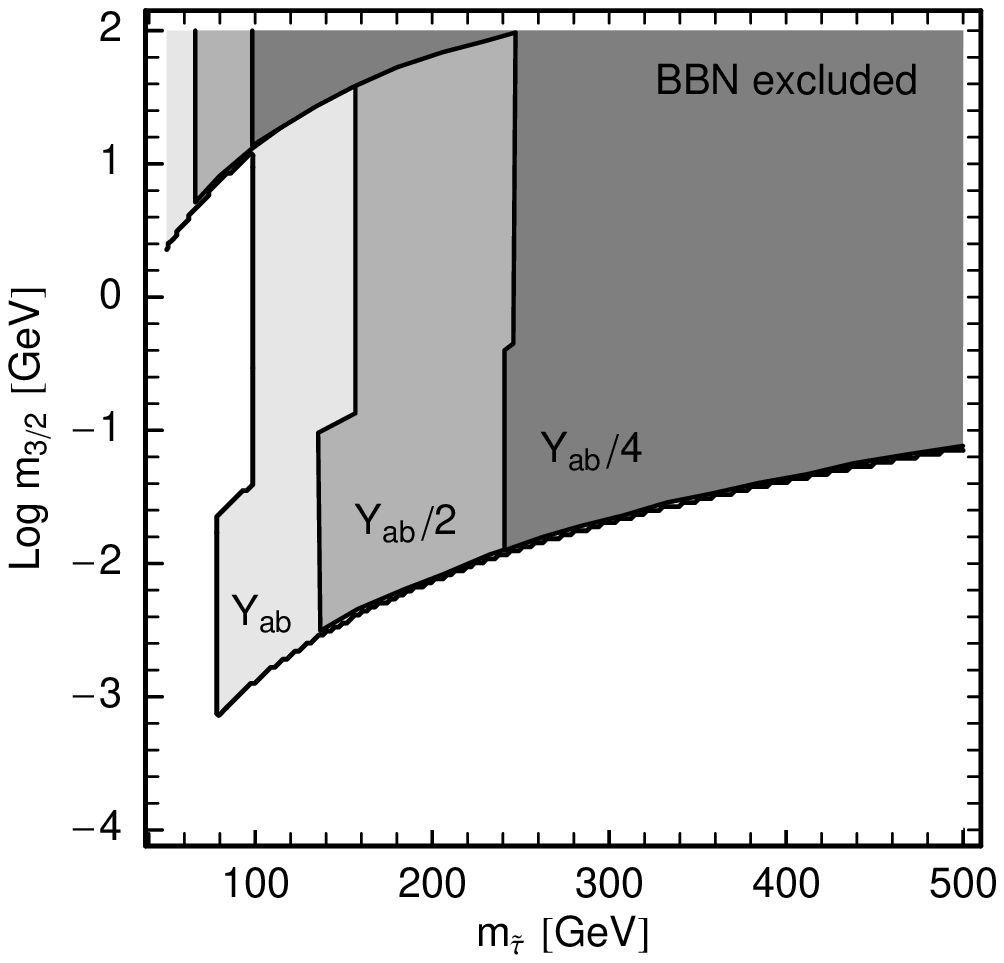}\quad
\includegraphics[height=6.8cm]{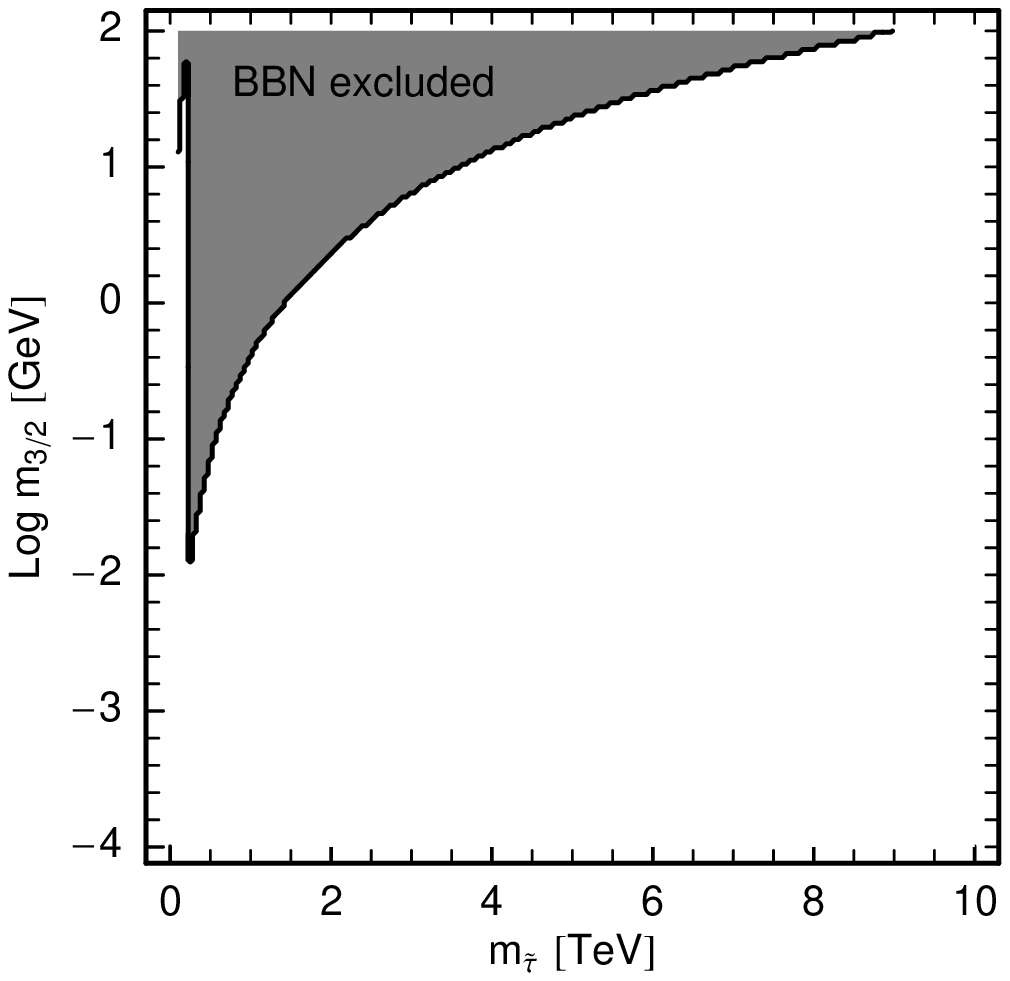}
\caption{\label{stau-gravitino}
BBN-excluded regions for a gravitino LSP in the plane $m_{3/2}$ vs.\ $m_{\tilde\tau} $.
On the left a zoom on $m_{\tilde\tau}=50$--$500$~GeV
for $Y_{\tilde\tau}= Y_{ab}$ (light grey ), $Y_{\tilde\tau}= Y_{ab}/2$ (medium grey)
and $Y_{\tilde\tau}= Y_{ab}/4$ (dark grey). On the right for $m_{\tilde\tau}=0.1$--$10$~TeV.
Note that LEP excluded $ m_{\tilde\tau} \leq 99.4\; \mbox{GeV} $ for a charged particle
stable within the detector \cite{LEP-stau}. }
\end{figure}

For a gravitino LSP, the decay rate is given by~\cite{bchrs05}
\beq
\Gamma (\tilde\tau_R\rightarrow \tau \tilde G)
= (5.9 \times 10^8\;\mbox{s} )^{-1}
\left({m_{\tilde\tau}\over 100\; \mbox{GeV}} \right)^5
\left({100\; \mbox{GeV}\over m_{3/2}} \right)^2
\left(1 - {m_{3/2}^2\over m_{\tilde\tau}^2} \right)^4\,,
\label{eq-WidthGravitino}
\eeq
which typically gives longer stau lifetimes than the axino case.
Figure~\ref{stau-gravitino} shows the BBN-excluded region
in the $m_{3/2} $ vs $m_{\tilde\tau} $ plane. We consider
a number density $Y_{\tilde\tau} $ equal to 1/2 and 1/4 times $Y_{ab}$ to
account for the possible variation depending on $m_{\tilde B}$.
As can be seen, to avoid all bounds we need either a very light gravitino
in the MeV range for $m_{\tilde\tau}\sim {\cal O}(100)$~GeV,
or a very heavy stau, 
e.g.\ $m_{\tilde\tau} \gsim 1.4$~TeV ($9$~TeV) for 
$m_{3/2} = 1$~GeV ($100$~GeV),
corresponding to a stau lifetime shorter than $0.1$\,s.
On the other hand, for $m_{\tilde\tau}\sim 100$--$250$~GeV and a lifetime longer than
$100$\,s, $B_H\sim 10^{-3}$ can bring the effective yield below the bound of
$mY\approx 5\times 10^{-14}$ required by hadronic showers.
Last but not least, note that the constraint from catalysed BBN becomes 
stronger than the hadronic ones for lifetimes longer than about $ 10^4$\,s 
and excludes a light stau NLSP for gravitino masses above 10-100 GeV.

\subsection{Relic stop}

To discuss the case of a relic stop, we assume that only $\tilde t_R$ is light
while all other SUSY particles are heavy and decouple. Moreover, we assume
that the light Higgs is SM-like with a mass of $m_h=115$~GeV, and that the
other Higgs bosons are also heavy and do not contribute to the stop annihilation.

\begin{figure}[t]\centering
\includegraphics[height=7.6cm]{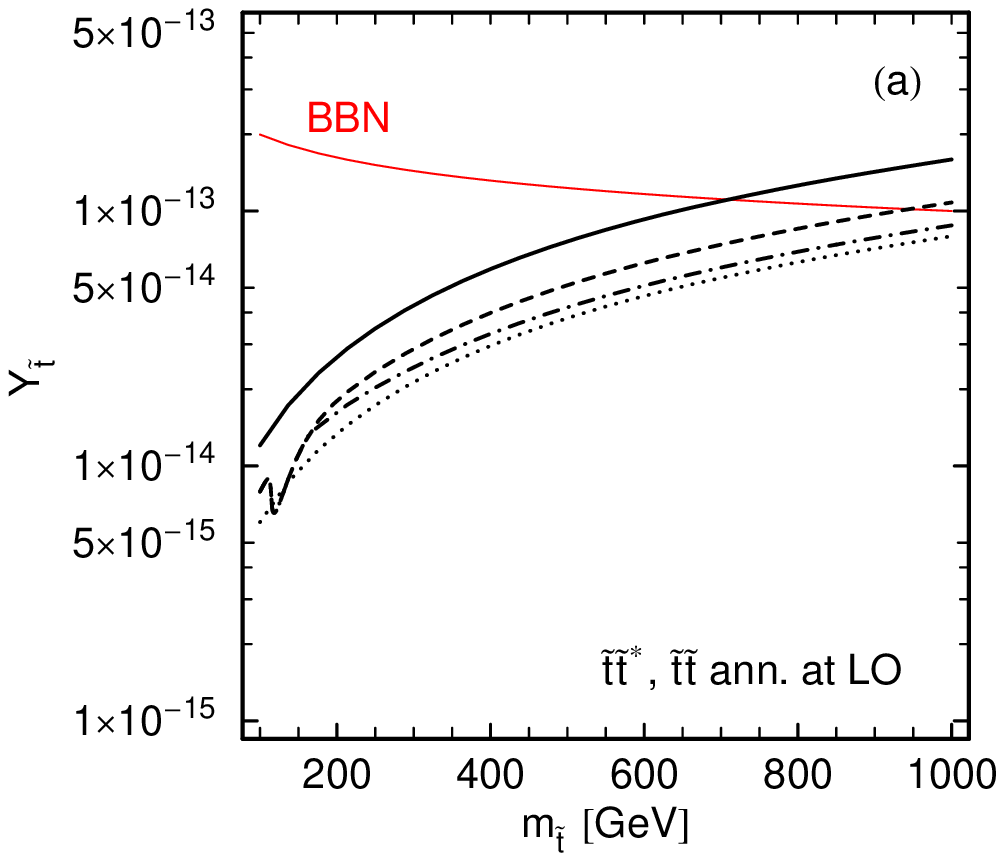}\quad
\includegraphics[height=7.6cm]{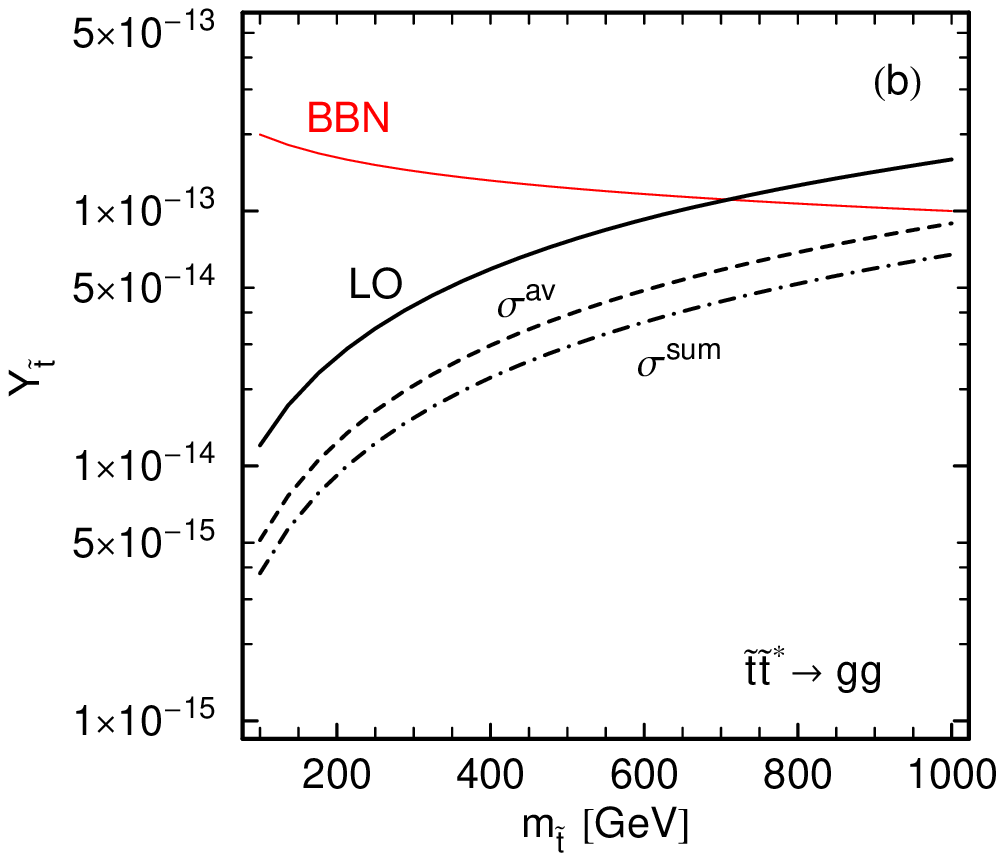}
\vspace*{-4mm}
\caption{\label{stop-yield}
Results for $Y_{\tilde t}$ for a relic $\tilde t_R$ as a function 
of the stop mass. 
In (a), tree-level results for different channels: the solid line comes 
from $\tilde t\tilde t^*\to gg$ only, the dashed line includes all 
channels into QCD+EW gauge and $h$ bosons (case of decoupled sparticles 
and heavy Higgses), the dash-dotted line is the result for 
$m_{\tilde g}=2m_{\tilde t_R}$, and the dotted line the limit 
$Y_{\tilde t}=Y_{nab}/2$.
In (b), the effect of the Sommerfeld enhancement on the yield from
$\tilde t\tilde t^*\to gg$: the full line shows the tree level result, 
the dashed line the result for $\sigma^{\mbox{\tiny SF} \rm av}$, 
i.e.\ applying an averaged Sommerfeld 
factor $C^{\rm av}_{SU(3)}=11/42$, and
the dash-dotted line is for $\sigma^{\mbox{\tiny SF} sum}$, 
i.e.\ applying a summed factor according to eq.~(\ref{SF-summed-SU3}).
The BBN bound for 0.1--100\,s lifetime is shown as thin red line 
in both plots.
}
\end{figure}

Results for the yield as a function of the stop mass are shown in
Fig.~\ref{stop-yield}. Let us first discuss the left plot,
Fig.~\ref{stop-yield}(a), which shows the yield at leading order
(LO). Here the full line is the pure QCD result, $Y_{nab}$ for
$SU(3)$, without Sommerfeld correction. As can be seen, $\tilde
t\tilde t^*\to gg$ alone is efficient enough to avoid the BBN
constraints up to stop masses of about 700 GeV. In the full
theory, the stop can also annihilate into other particles, in
particular into EW gauge and Higgs bosons. The yield for the
QCD+EW case, still assuming heavy sparticles, is shown as the dashed
line in Fig.~\ref{stop-yield}(a). The dip at $m_{\tilde t}\sim
120$ GeV is due to the onset of $\tilde t_R^{}\tilde t_R^*\to hh$. Other important
channels are annihilation into $W^+W^-$ and $\gamma g$,
contributing about 10\% each to the total annihilation cross
section for $m_{\tilde t}\gsim 200$~GeV. Annihilation into $ZZ$
contributes about 5\%. Annihilation into top quarks is suppressed
by the heavy gluino mass, and also by $m_t$. However, if
$m_{\tilde t}>200$~GeV and $m_{\tilde g}\sim 2m_{\tilde t}$,
$\tilde t_R\tilde t_R\to tt$ further reduces the yield by
10--20\%. This is shown as the dash-dotted line in
Fig.~\ref{stop-yield}(a). All in all, annihilation into gluons is,
however, always the dominant channel, contributing at least 50\%.
We therefore take $Y_{nab}/2$ as a rough limit, which is shown as the
dotted line in Fig.~\ref{stop-yield}(a). Comparing with the BBN
constraints we see that a relic $\tilde t_R$ with a lifetime of
0.1--100\,s can be in agreement with BBN even for high masses of
about 1 TeV.

The impact of the Sommerfeld enhancement is illustrated in 
Fig.~\ref{stop-yield}(b) for the case $\tilde t\tilde t^*\to gg$. 
As can be seen, taking the averaged Sommerfeld factor of 
$C^{av}_{SU(3)}=11/42$ in eq.~(\ref{singlechannel}) reduces 
the LO yield by roughly a factor of 2, while a summed factor 
according to eq.~(\ref{SF-summed-SU3}) reduces the LO yield by 
roughly a factor of 3. These results are in qualitative agreement
with those of \cite{Freitas07}, that considered the Sommerfeld 
correction in the neutralino-stop coannihilation region.
Here note that for colour-singlet channels like, for instance, 
$\tilde t\tilde t^*\to W^+W^-$
a factor of $C=4/3$ applies, hence leading to even larger enhancement.
We leave a detailed numerical analysis of the enhancement of the various stop annihilation
channels for future work. Here we just note that the overall effect can be a reduction of the
yield by an order of magnitude.

Additional annihilation can take place after the QCD phase transition, when
the stops are in a confined phase with the quarks. Since the lighter fermionic
mesons are neutral and assuming that the annihilation process takes place
without the formation of a bound state between the mesinos, the unitarity
cross section is probably a good estimate of such annihilation and allows
for heavier stops to be consistent with hadronic shower constraints.
We see in fact from Fig.~\ref{Yofm} that the unitary cross section
with three degrees of freedom gives a yield well below all the BBN 
bounds (and below the range in Fig.~\ref{stop-yield}) for stop masses
up to 700 GeV. If also bound states between the mesinos can form efficiently, 
the BBN constraints disappear altogether~\cite{Kang:2006yd}, but note that 
we do not have to rely on the enhancement coming from such processes, 
which are very difficult to compute, for a wide range of parameter space.

Let us briefly discuss the lifetime also for the stop case.
For the case of an axino LSP, the stop decay rate is
a larger than for the stau since it depends on the
gluino mass and the QCD gauge coupling \cite{crs02}:
\beq
\Gamma (\tilde t_R\rightarrow t \tilde a)
= (1.3 \times 10^{-3}\; \mbox{sec})^{-1} \xi_t^2
\left({m_{\tilde t}\over 10^2\; \mbox{GeV}} \right)
\left({m_{\tilde g}\over 10^2\; \mbox{GeV}} \right)^2
\left({10^{11}\; \mbox{GeV}\over f_a} \right)^2
\left(1 - {m_{\tilde a}^2\over m_{\tilde t}^2} \right)
\eeq
where $\xi_t $ is again a factor of order one taking into account
the uncertainties in the loop computation \cite{bchrs05},
in principle different than the one for the stau.
Therefore, for the axino case, the BBN bound never applies
if the decay into top is kinematically allowed, i.e.\ if
$m_{\tilde t}^2 \geq (m_{\tilde a} + m_{t})^2$.
If the stop mass is smaller, the decay can proceed through
a virtual top, for which we estimate a suppression of order
${\cal O} (1/100) $ due to the 3-body phase space.
This would still give a lifetime of order 0.1 sec, so
the BBN constraints are completely avoided,
as long as there is not a strong degeneracy in mass
between LSP and NLSP or the factor $\xi_t $ is
exceptionally large.

\begin{figure}[t]\centering
\includegraphics[width=7cm]{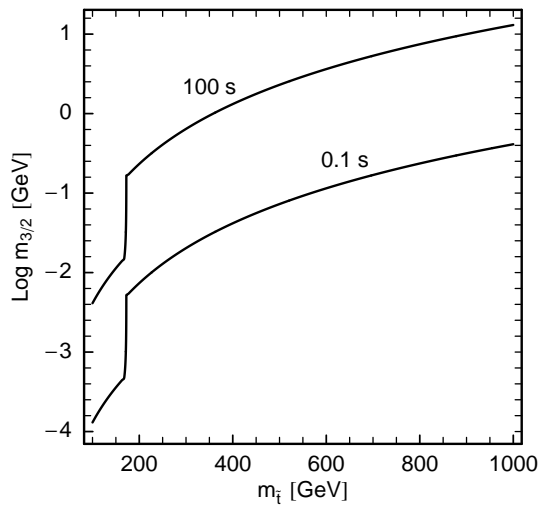}
\caption{\label{stop-gravitino}
Lifetime of a relic $\tilde t_R$ in the plane $m_{3/2}$ vs $m_{\tilde t} $.
Recent results from the Tevatron exclude a metastable
stop below 250 GeV \cite{TeVatron-stop}.
}
\end{figure}

For a gravitino LSP, on the other hand, the same formula
applies for stop as for stau, eq.~(\ref{eq-WidthGravitino})
with $\tilde\tau\to\tilde t$, because the gravitino couples
only to mass. Note, however, that also
in this decay the width gets phase-space suppressed if
$m_{\tilde t}<m_t+m_{\tilde G}$. For illustration, we show
in Fig.~\ref{stop-gravitino} the band of 0.1--100\,s lifetime
in the plane $m_{3/2}$ vs $m_{\tilde t}$.
For lifetimes longer than 100 s, stops can still be in accord
with BBN thanks to the additional annihilation during the QCD 
phase transition, if their annihilation reaches the unitarity one.
We therefore conclude that cosmologically stops are an allowed 
NLSP in any mass range and in particular also for a heavy gravitino.
Our results are in agreement with those for specific supersymmetric 
models with stop NLSP discussed in \cite{stop-gravitino}.
From the colliders side, note that the low mass region $m_{\tilde t}<250$~GeV 
has been recently excluded by the search for charged massive particles 
at the Tevatron~\cite{TeVatron-stop}.

\section{Conclusions}

We have studied the number density of a charged relic by computing the
annihilation cross section into gauge bosons, including the Sommerfeld
enhancement. We have found that the Sommerfeld factor increases the
thermally averaged annihilation cross section by 20-50\% and reduces 
the final yield even by a factor 2 or 3 for the $SU(3)$ case. 
Moreover the result is very sensitive on how the higher orders are resummed. 
Nevertheless the number density surviving the annihilation is still large 
and BBN constraints are relevant for most relics. They can be avoided 
completely only for very large $N$ for particles in the fundamental 
representation of $SU(N)$
($N> 100 $ for $ m_X \leq 10 $ TeV) or for cross sections nearly
fulfilling the unitarity bound.
For the cases of SM gauge groups, the allowed regions only correspond to
very  light relic masses, where the number density is low enough, or
to sufficiently heavy relic masses so that the decay takes place in the
first stages of BBN.
The latter allowed region depends strongly on the relic decay channel,
and, in case of a gravitino LSP with conserved R-parity,
also on the gravitino mass. 
Let us mention here that if R-parity is just marginally
broken, the NLSP can decay with shorter lifetime through R-parity
violating channels and the BBN constraints can be easily evaded for
any NLSP while keeping the gravitino LSP as Dark Matter~\cite{R-parity}.

More specifically, for the stau NLSP the light mass window has nearly
completely been excluded by direct searches at LEP, even if the
annihilation cross-section is maximal
$ \sim 4 \sigma(\tilde\tau\tilde\tau^* \rightarrow \gamma\gamma) $,
unless the gravitino is lighter than a few tens of GeV,
while the large mass region is unfortunately out of reach at the
LHC for gravitino masses $m_{3/2} > 100 \mbox{GeV} $.
The detection of a quasi-stable stau at the LHC would then point
to a scenario with relatively light gravitino mass, R-parity
breaking or an axino LSP and could probably exclude the gravity mediated 
supersymmetry breaking scenario. In that case the determination
of the stau lifetime and its decays will become crucial in distinguishing
the different LSPs \cite{stauatcolliders,bchrs05}.

The stop case is much less constrained thanks to the stronger annihilation
cross-section, even if in this case the decay always produces mainly 
hadrons.
We have practically no constraints if the LSP is an axino and
even for a gravitino LSP, we can allow for relatively light stops 
up to approximately 700 GeV ($1\;$ TeV for lifetimes below 
$10^{7}\;$ s), if the annihilation cross section reaches the unitarity 
one after the QCD phase transition. 
The window between the present Tevatron bound around 250 GeV and 1 TeV 
should be surely completely covered by the LHC, the signature being a 
quasi-stable heavy fermionic meson.
The detection of such a state would call for a non-minimal SUSY breaking 
sector with a coloured NLSP and a very weakly interacting LSP.
In this case again only the analysis of the stop decays would allow
to distinguish between the lightest states.

\section*{Acknowledgements}
\addcontentsline{toc}{section}{Acknowledgements}

We thank Nans Baro, Fawzi Boudjema, Lance Dixon, Koichi Hamaguchi,
Wolfgang Kilian, Tom Rizzo, Masato Senami and Iain Stewart for 
useful discussions and communications.
We thank specially Alexander Pukhov for valuable help with implementing
the Sommerfeld enhancement in Micromegas.

CFB's work is supported in part by funds provided by the 
U.S. Department of Energy (D.O.E.) under cooperative research 
agreement DE-FC02-94ER40818 and by the National Science
Foundation under Grant No.\ PHY05-51164.

LC would like to thank the Institute of Theoretical Physics of the 
Warsaw University for their hospitality during part of this work; 
the visit was supported by a Maria Curie Transfer of Knowledge Fellowship 
of the European Community's Sixth
Framework Programme under contract number MTKD-CT-2005-029466 (2006-2010).
LC also acknowledges the support of the "Impuls- und Vernetzungsfond" of the
Helmholtz Association under the contract number VH-NG-006 and of the European
Network of Theoretical Astroparticle Physics ILIAS/N6 under contract number
RII3-CT-2004-506222.

This work is also part of the French ANR project ToolsDMColl, BLAN07-2-194882.


\appendix 

\section{Annihilation into massless $SU(N)$ gauge bosons}

\subsection{Amplitudes for the annihilation}

We consider the case of one particle and antiparticle in the representation
$ T^a_i$ and its conjugate, with momenta $p_1,p_2$ and mass $m$ annihilating
into two massless gauge bosons with group indices $a,b$, momenta $p_3, p_4$
and Lorentz indices $\mu,\nu$ respectively.

The process has four different contributions,
corresponding to the following four Feynman diagrams:
\begin{itemize}
\item[\bf t]{particle exchange in the t-channel described
by the amplitude
\beq
{\cal A}_t^{\mu\nu} =
i g_N^2 \left(T^b T^a\right)_{ji}
{(2 p_1-p_3)^{\mu} (2p_2-p_4)^\nu \over t-m^2} \, ;
\eeq
}
\item[\bf u]{particle exchange in the u-channel described
by the amplitude
\beq
{\cal A}_u^{\mu\nu} =
i g_N^2 \left(T^a T^b\right)_{ji}
{(2 p_1-p_4)^{\nu} (2p_2-p_3)^\mu \over u-m^2} \, ;
\eeq
note that this contribution is identical to the t-channel
under interchange of $a \leftrightarrow b$, $ \mu  \leftrightarrow \nu$,
$ (p_3, t )  \leftrightarrow ( p_4, u ) $;
}
\item[\bf 4]{supersymmetric four-scalar coupling giving the
amplitude:
\beq
{\cal A}_4^{\mu\nu} =
i g_N^2 \left\{ T^a, T^b \right\}_{ji} g^{\mu\nu} \, ;
\eeq
this contribution is symmetric in the exchange of $a,b$ and therefore
also $\mu,\nu$;
}
\item[\bf s]{off-shell gauge boson in the s-channel decaying into
two bosons via the non-abelian interaction:
\bea
{\cal A}_s^{\mu\nu} &=&
- i g_N^2 \left[ T^a, T^b \right]_{ji} {1\over s}
\left[ g^{\mu\nu} (t-u) - (2p_4+p_3)^{\mu} (p_1-p_2)^{\nu}
\right.\nonumber\\
& &\left.
+ (p_1-p_2)^{\mu} (2p_3+p_4)^{\nu} \right] ;
\eea
this contribution is completely antisymmetric under the exchange
of the gauge bosons group indices and therefore also under the
exchange of their momenta and Lorentz indices.
}
\end{itemize}

For convenience, we can then separate the amplitude into
symmetric and antisymmetric part in colours $a,b$; then
the interference between the two parts vanishes. Using
\beq
T^a T^b = {1\over 2}  \left\{ T^a, T^b \right\} +
{1\over 2}  \left[ T^a, T^b \right]
\eeq
we have then
\bea
{\cal A}_{sym}^{\mu\nu} &=&
\frac{i g_N^2}{2} \left\{T^a, T^b\right\}_{ji}
\left[ \frac{(2 p_1-p_3)^{\mu} (2p_2-p_4)^{\nu} }{ t-m^2}
+ \frac{(2 p_1-p_4)^{\nu} (2p_2-p_3)^{\mu} }{ u-m^2}
+ 2 g^{\mu\nu} \right]
\nonumber\\
\eea
and
\bea
{\cal A}_{asym}^{\mu\nu} &=&
\frac{i g_N^2}{2} \left[T^a, T^b\right]_{ji}
\left[ - \frac{(2 p_1-p_3)^{\mu} (2p_2-p_4)^{\nu} }{ t-m^2}
+ \frac{(2 p_1-p_4)^{\nu} (2p_2-p_3)^{\mu} }{ u-m^2}
\right. \nonumber\\
& & \left.
- 2 \frac{ g^{\mu\nu} (t-u) - (2p_4+p_3)^{\mu} (p_1-p_2)^{\nu} +
(p_1-p_2)^{\mu} (2p_3+p_4)^{\nu} }{ s}
\right] .
\eea
In the Boltzmann equation, we have to insert the averaged cross-section,
so first we have to sum over all the final  and initial states, i.e.  sum over
the gauge bosons polarisations and over all the group indices.

\subsection{The matrix element}

The computation for the symmetric piece is straightforward:
\bea
|{\cal A}_{sym} |^2 &=&
g_N^4  |\left\{T^a, T^b\right\}_{ji}|^2
\left[
{(t+m^2)^2 \over (t-m^2)^2} + {(u+m^2)^2 \over (u-m^2)^2}
+ {1\over 2} {(s-4m^2)^2 \over (t-m^2)(u-m^2)}
\right.\nonumber\\
& & \left.
+ 4 + {s/2-4m^2- 2(t-m^2) \over t-m^2} +  {s/2-4m^2- 2(u-m^2) \over u-m^2}
\right]\\
&=&
4 g_N^4  |\left\{T^a, T^b\right\}_{ji}|^2
\left[ {1\over 2}  + {2 m^4 \over (t-m^2)^2}
+ {2 m^2\over t-m^2} \left(1- {2 m^2\over s}  \right)
\right] .
\eea

In the antisymmetric part instead  we have to take into account ghost subtraction
and the total result is
\bea
|{\cal A}_{asym} |^2 &=&
g_N^4  |\left[T^a, T^b\right]_{ji}|^2
\left[
{(t+m^2)^2 \over (t-m^2)^2} + {(u+m^2)^2 \over (u-m^2)^2}
- 4
\right.\nonumber\\
& & \left.
- {1\over 2} {(s-4m^2)^2 \over (t-m^2)(u-m^2)}
+ 2 {(t-u)^2 \over s^2} + {16 m^2\over s}
\right.\nonumber\\
& & \left.
+ {(t-u)(3/2 s - t -3m^2)+ 2(s-4m^2) (u-m^2) \over s (t-m^2)}
\right. \nonumber\\
& & \left.
+  {(u-t)(3/2 s - u -3m^2)+ 2(s-4m^2) (t-m^2) \over s (u-m^2)}
\right]
\label{antisym2a}\\
&=&
4 g_N^4  |\left[T^a, T^b\right]_{ji}|^2
\left[
  {(t-u)^2 \over 2 s^2} + {4 m^2\over s}
+ {2 m^4 \over (t-m^2)^2}
\right.\nonumber\\
& & \left.
+ {2 m^2 \over t-m^2} \left( 1 + {2 m^2\over s}\right)
\right] .
\eea

So for the total matrix element  we have
\bea
|{\cal M}|^2 &=& 4 g_N^4 \left\{
|\left\{T^a, T^b \right\}_{ji}|^2
\[{1\over 2} + { 2 m_X^4 \over (t-m_X^2)^2} + {2m_X^2 \over t-m_X^2}
\left( 1- {2m_X^2\over s} \right)\right]
\right. \nonumber\\
& & \left. + |\left[T^a, T^b \right]_{ji}|^2
\left[ {1\over 2} {(s + 2 (t-m_X^2))^2 \over s^2}
+ {4 m_X^2 \over s}
+ { 2 m_X^4 \over (t-m_X^2)^2}
\right.\right. \nonumber\\
& & \left. \left.
+ {2m_X^2 \over t-m_X^2}
\left( 1+ {2m_X^2\over s} \right)\right] \right\} \; .
\eea
and the cross section is given in eq.~(\ref{sigmaN}).

\subsection{Comparison with QCD result}

For the case of $SU(3) $ we have

\bea
\sum_{a,b,i,j}\, \Big|\left\{T^a, T^b\right\}_{ji}\Big|^2 &=& {28\over 3}
\eea
and
\bea
\sum_{a,b,i,j}\,\Big|\left[T^a, T^b\right]_{ji}\Big|^2 &=&
{1\over 2} \sum_{a,b,c} f_{abc}^2 = 12\, .
\eea

So after the sum over colours, we get
\bea
|{\cal M}|^2 &=& 4 g_3^4
\left[ {14\over 3} + 6 {(t-u)^2 \over s^2} + {48 m^2\over s}
+ {2 m^4 \over (t-m^2)^2} \left( {28\over 3} + 12 \right)
\right.\nonumber\\
& & \left.
+ {2 m^2 \over t-m^2} \left( {28\over 3} + 12
+ {2 m^2\over s} \left(- {28\over 3} + 12\right) \right)
\right]\\
&=&
 4 g_3^4
\left[ {32\over 3} + 24 { t-m^2 \over s} + 24 {(t-m^2)^2 \over s^2}
+ {48 m^2\over s}
\right.\nonumber\\
& & \left.
+ {128 \over 3} { m^4 \over (t-m^2)^2}
+ {128\over 3} {m^2 \over t-m^2} \left( 1
+ {1 \over 4} {m^2\over s} \right)
\right]\; .
\label{M-QCD}
\eea

This result coincides with the one given in the literature for the
QCD case~\cite{bhsz96}.
Compare in general with \cite{bhsz96}:
\bea
|{\cal M} (gg\rightarrow \tilde q \bar{\tilde q})|^2
&=& 4 n_f g_3^4 \left[ C_0 \left(1-2 {(t-m^2) (u-m^2)\over s^2 } \right)
-C_K \right] \times \\
& & \times \left[1-2 {s m^2 \over (t-m^2) (u-m^2) }
\left( 1 -   {s m^2 \over (t-m^2) (u-m^2) } \right) \right]
\nonumber\\
&=& 4 n_f g_3^4 \left[
C_0 -C_K + 2 C_0 {t-m^2 \over s} +  2 C_0 {(t-m^2)^2 \over s^2}
\right.\nonumber\\
& & \left.
+ 4 C_0 {m^2 \over s} + 4 (C_0 -C_K) {m^4 \over (t-m^2)^2}
\right.\nonumber\\
& & \left.
+ 4 {m^2\over t-m^2} \left(C_0-C_K+ 2 C_K {m^2\over s}\right)
\right]
\eea
using again the symmetry in $u \leftrightarrow t$ and eliminating $u$.

We have also that
\beq
C_0 = \sum_{a,b,c} f_{abc}^2 = N (N^2-1) = 24 \quad\quad
C_K = {N^2-1\over N} = {8\over 3}
\eeq
and for a single RH stop, we must use $ 2 n_f = 1$. Then we get
\bea
|{\cal M} (gg\rightarrow \tilde t_{R} \bar{\tilde t_{R}})|^2
&=& 4 g_3^4 \left[
{32 \over 3} + 24 {t-m^2 \over s} + 24 {(t-m^2)^2 \over s^2}
+ 48 {m^2 \over s} \right.\nonumber\\
& & \left.
+ {128\over 3} {m^4 \over (t-m^2)^2}
+ {128\over 3} {m^2\over t-m^2} \left( 1 + {1\over 4} {m^2\over s}\right)
\right]\; ,
\eea
which coincide with our result above eq.~(\ref{M-QCD}).

Now integrate over $t$ and obtain
\bea
\sigma (m, s) &=& 32\; {4 \pi \alpha_3^2 \over s - 4m^2}
\left[
\sqrt{1-{4m^2\over s} } \left( {5\over 24}
+ {31 \over 12} { m^2 \over s}\right)
\right.\nonumber\\
& & \left. +  {4\over 3} {m^2\over s}
\left(1 + {1\over 4} {m^2\over s}\right)
\log \left({1-\sqrt{1-{4m^2\over s}}\over
1+ \sqrt{1-{4m^2\over s}}} \right)
\right]\; ,
\eea
which coincides with \cite{bhsz96} allowing for the
exchange of initial and final state ($ s - 4 m^2 \rightarrow s$
in the denominator) and the initial state averaging, i.e.
a factor of 1/64 for the two gluons initial state.

\section{Annihilation into $SU(2)_L $ gauge bosons}

Another important channel of annihilation for light stops or staus
is into EW gauge bosons. Let us consider first the pure $ SU(2)_L$ case,
neglecting the gauge boson masses, but with a split  $ SU(2) $ multiplet.
We consider here the case of one left-handed sparticle and one 
left-handed antisparticle
of momenta $p_1,p_2$, mass $m_1$ and $SU(2) $ index $1$,
annihilating into 2 gauge bosons  of $ SU(2)_L$ index $i,j$,
momenta $p_3, p_4$ and Lorentz indices $\mu,\nu$ respectively.
Then we can directly use the result for $SU(N)$, only taking into account
that $T^a \rightarrow \sigma^i/2 $, with $  \sigma^i $
denoting the Pauli matrices, and that in this case we have
an initial state made of the upper components of the
$ SU(2)_L$ doublet, while the lower component is exchanged in
the t- and u-channel and can have a different mass $m_2 $.

We have then for the two amplitudes, symmetric and antisymmetric
in the group and Lorentz indices,
\bea
{\cal A}_{sym}^{\mu\nu} &=&
i \frac{g_2^2}{8} \left\{\sigma^i, \sigma^j\right\}_{11}
\left[  \frac{(2 p_1-p_3)^{\mu} (2p_2-p_4)^{\nu} }{ t-m_2^2}
+ \frac{(2 p_1-p_4)^{\nu} (2p_2-p_3)^{\mu} }{ u-m_2^2}
+ 2 g^{\mu\nu} \right]
\nonumber\\
\eea
and
\bea
{\cal A}_{asym}^{\mu\nu} &=&
i \frac{g_2^2}{8} [\sigma^i, \sigma^j]_{11}
\left[ - \frac{(2 p_1-p_3)^{\mu} (2p_2-p_4)^{\nu} }{ t-m_2^2}
+ \frac{(2 p_1-p_4)^{\nu} (2p_2-p_3)^{\mu} }{ u-m_2^2}
\right.\nonumber\\
& & \left.
-  2 \frac{ g^{\mu\nu} (t-u) - (2p_4+p_3)^{\mu} (p_1-p_2)^{\nu} +
(p_1-p_2)^{\mu} (2p_3+p_4)^{\nu} }{ s}
\right] .
\eea
To compute the annihilation cross section, we have to sum
over all the final states and initial states; this means that we have to sum over the W
polarisations and over the $ SU(2)_L $ indices $i,j$, but in this case the
initial state group indices are fixed.

The symmetric piece gives
\bea
|{\cal A}_{sym} |^2 &=&
\frac{g_2^4}{16}  | \left\{\sigma^i, \sigma^j\right\}_{11} |^2
\left[
{(t+m_1^2)^2 \over (t-m_2^2)^2} + {(u+m_1^2)^2 \over (u-m_2^2)^2}
+ {1\over 2} {(s-4m_1^2)^2 \over (t-m_2^2)(u-m_2^2)}
\right.\nonumber\\
& & \left.
+ 4 + {s/2-4m_1^2- 2(t-m_1^2) \over t-m_2^2} +  {s/2-4m_1^2- 2(u-m_1^2) \over u-m_2^2}
\right]\\
&=&
\frac{g_2^4}{4}  | \left\{\sigma^i, \sigma^j\right\}_{11} |^2
\left[ \frac{1}{2}  +  \frac{1}{2} {(m_1^2+m_2^2)^2 \over (t-m_2^2)^2}
\right.\\
& & \left.
+ {1\over t-m_2^2} \left(
\frac{3m_2^2 + m_1^2}{2} -
\frac{(m_1^2 + m_2^2)^2 }{s+ 2m_2^2 - 2m_1^2}  \right)
\right] .
\nonumber
\eea

In the antisymmetric part instead  gives
\bea
|{\cal A}_{asym} |^2 &=&
\frac{g_2^4}{16} | [\sigma^i, \sigma^j]_{11} |^2
\left[
{(t+m_1^2)^2 \over (t-m_2^2)^2} + {(u+m_1^2)^2 \over (u-m_2^2)^2}
- 4
\right.\\
& & \left.
- {1\over 2} {(s-4m_1^2)^2 \over (t-m_2^2)(u-m_2^2)}
+ 2 {(t-u)^2 \over s^2} + {16 m_1^2\over s}
\right.\nonumber\\
& & \left.
+ {(t-u)(3/2 s - t -3m_1^2)+ 2(s-4m_1^2) (u-m_1^2) \over s (t-m_2^2)}
\right. \nonumber\\
& & \left.
+  {(u-t)(3/2 s - u -3m_1^2)+ 2(s-4m_1^2) (t-m_1^2) \over s (u-m_2^2)}
\right]
\nonumber\\
&=&
\frac{g_2^4}{4} | [\sigma^i, \sigma^j]_{11} |^2
\left[
\frac {(t-u)^2 }{  2 s^2} +
\frac{5 m_1^2 - m_2^2}{s}
+ \frac{1}{2} \frac{(m_1^2+m_2^2)^2 }{(t-m_2^2)^2}
\right.\\
& & \left.
+ {1 \over t-m_2^2}
\left( \frac{m_2^2 + 3 m_1^2}{2}
+ \frac{(m_2^2 +  m_1^2)^2 }{s+ 2m_2^2 - 2m_1^2}
-  \frac{(m_2^2 - m_1^2)^2 }{ s}\right)
\right] .
\nonumber
\eea

\subsection{$ SU(2)_L $ sum and total matrix element}

In this case the sum over the indices $i,j$ is simple.
We have that
\bea
\sum_{i,j} \frac{1}{4} | \left\{\sigma^i, \sigma^j\right\}_{11}|^2 &=&
\sum_{i,j}  \frac{1}{4}  | 2 \delta_i^j I_{11}|^2
= \sum_i  \delta_i^i = 2 + 1
\eea
where we have considered the annihilation into
$W^{1,2} $ separately from that into $W_3$.
In fact the intermediate particle has a different mass in the two cases.

On the other hand the antisymmetric product gives
\bea
\sum_{i,j} \frac{1}{4}  |[\sigma^i, \sigma^j]_{11}|^2 &=&
\sum_{i,j}  \frac{1}{4} | 2 \epsilon_{ijk} \sigma^k_{11} |^2 = \sum_{i,j} | \epsilon_{ij3} |^2 = 2
\eea
since in this case only $W^3 $ can be exchanged in the s-channel
for $W^{1,2} $ in the final state.

Then the matrix element for annihilation into $W^{1,2}$ gauge
bosons is given by
\bea
|{\cal M}_{W12} |^2 &=&
g_2^4
\left[ 1  +
\frac {(t-u)^2 }{s^2} + \frac{10 m_1^2 - 2 m_2^2}{s}
+ 2  {(m_1^2+m_2^2)^2 \over (t-m_2^2)^2}
\right.\\
& & \left.
+  \frac{4 }{ t-m_2^2}
\left(m_2^2 + m_1^2
- \frac{(m_2^2 - m_1^2)^2 }{ 2 s}\right)
\right] \; ,
\nonumber
\eea
while the annihilation into $ W^3 $ has only the abelian contribution
with the presence of a single mass $m_1$
\bea
|{\cal M}_{W3} |^2 &=&
g_2^4
\left[ \frac{1}{2}  +  {2 m_1^4 \over (t-m_1^2)^2}
+ {2 m_1^2\over t-m_1^2} \left(
1-  \frac{ 2 m_1^2}{s}  \right)
\right] .
\nonumber
\eea

The cross section for the first case is then
\bea
\sigma_{W12} &=&
\frac{2\pi\alpha_2^2 }{ s - 4m_1^2}
\left[ \sqrt{1 - \frac{4 m_1^2}{s}}
\left(
\frac{2}{3} + \frac{13}{3} \frac{ m_1^2}{s}
- \frac{m_2^2}{s}   + \frac{(m_1^2+m_2^2)^2 }{s m_2^2 + (m_2^2-m_1^2)^2}
\right)
\right.\nonumber\\
& & \left.
+  2 \left( \frac{m_2^2 + m_1^2}{s}
- \frac{(m_2^2 - m_1^2)^2 }{ 2 s^2}\right) \times
\right.\nonumber\\
& & \left.
\times \log \left( \frac{s + 2 (m_2^2 - m_1^2) - \sqrt{s(s- 4m_1^2)} }{
s + 2 (m_2^2 - m_1^2) + \sqrt{s(s- 4m_1^2)}} \right)
\right] \; ,
\eea
while the annihilation into $ W^3 $ is identical to the abelian one
in eq.~(\ref{sigmaab}) for $e_X = 1/2 $.

\subsection{Annihilation into physical $W^+ W^-$}

Let us now consider the case of a broken $SU(2)_L$ symmetry
like the Standard Model and massive gauge bosons which mix
to give the physical $W^+,W^-,Z, \gamma $. At the same
time let us consider a general initial state given by
the light stau mass eigenstate
$\tilde \tau_1 = \tilde \tau_L \cos\theta_{\tilde \tau} +
\tilde \tau_R \sin\theta_{\tilde \tau} $
and its antiparticle. In this case the intermediate particle
exchanged in the t- and u-channel can be only a left-handed
sneutrino and therefore we can neglect the mixing for
the intermediate state.

Then the annihilation into $W^+ W^-$ is given by the
following channels:
\begin{itemize}
\item[\bf t]{sneutrino exchange in the t-channel described
by the amplitude
\beq
{\cal A}_t^{\mu\nu} =
i \frac{g_2^2}{2} \cos^2\theta_{\tilde \tau}
{(2 p_1-p_3)^{\mu} (2p_2-p_4)^\nu \over t-m_{\tilde \nu}^2} \, ;
\eeq
}
\item[\bf u]{NO u-channel since $W^+ $ and $W^-$ are
different particles !}

\item[\bf 4]{supersymmetric four-scalar coupling giving the
amplitude:
\beq
{\cal A}_4^{\mu\nu} =
i \frac{g_2^2}{2} \cos^2\theta_{\tilde \tau} g^{\mu\nu} \, ;
\eeq
this contribution is symmetric in the exchange of $\mu,\nu$;
}

\item[\bf s]{off-shell $Z/\gamma $ in the s-channel decaying
into two $WW$ via the non-abelian interaction:
\bea
{\cal A}_s^{\mu\nu}
&=&
i \frac{g_2^2}{2} \cos^2\theta_{\tilde \tau}
\left( 1 - \frac{4}{3} \frac{\sin^2\theta_W}{\cos^2\theta_{\tilde t}}\right)
\frac{1}{ s - M_Z^2 }
\left[ g^{\mu\nu} (t-u)
\right.\\
 & & \left.
- (2p_4+p_3)^{\mu} (p_1-p_2)^{\nu}
+ (p_1-p_2)^{\mu} (2p_3+p_4)^{\nu} \right] \nonumber\\
& &
+ i e^2 \frac{2}{3} \frac{1}{s}
\left[ g^{\mu\nu} (t-u) - (2p_4+p_3)^{\mu} (p_1-p_2)^{\nu}
\right.  \nonumber\\
& & \left.
+ (p_1-p_2)^{\mu} (2p_3+p_4)^{\nu} \right] \nonumber\\
&=&
i \frac{g_2^2}{2} \cos^2\theta_{\tilde \tau}
\left( 1- \frac{4}{3} \frac{\sin^2\theta_W}{\cos^2\theta_{\tilde \tau}}
\frac{M_Z^2}{s} \right) \\
& &
\frac{g^{\mu\nu} (t-u)- (2p_4+p_3)^{\mu} (p_1-p_2)^{\nu}
+ (p_1-p_2)^{\mu} (2p_3+p_4)^{\nu}}{ s - M_Z^2} \nonumber
\; ;
\eea
this contribution is completely antisymmetric under the exchange
of the W momenta and Lorentz indices.
Note that the photon contribution is proportional to
$ e^2 = g_2^2 \sin^2\theta_W $ and cancels exactly with the
second term due to the Z-boson in the case of equal mass.
In that limit in fact the $U(1)_Y $ factor decouples and
does not participate in the non-abelian interaction.
}

\item[\bf s-H]{off-shell $h/H $ in the s-channel decaying
into two $WW$ via the non-abelian interaction;
in this case we have to consider both neutral Higgses:
\bea
{\cal A}_{sH}^{\mu\nu}
&=&
i \frac{g_2^2}{2} \cos^2\theta_{\tilde \tau} g^{\mu\nu} \frac{M_W^2}{s}
\left[
\frac{C_H s}{ s - M_H^2 } + \frac{C_h s}{ s - M_h^2 }\right]
\; ;
\eea
where $C_{H/h} $ is coming from the product of the coupling
of the staus to the Higgses and of the Higgses to the WW pair.
These constants depend on the whole SUSY breaking parameters.
For the staus these couplings are probably negligible.
We have in fact
\bea
C_{H/h} &=&
\frac{(Z_{1H/h})^2 - (Z_{2H/h})^2 \tan^2\beta}{
(1+\tan^2\beta)\cos^4\theta_W}
\left(1 - \frac{4}{3} \sin^2\theta_W (1-\tan^2\theta_{\tilde \tau})
\right)
\nonumber\\
& & + 4\; \frac{Y_{\tau}^2 \tan\beta Z_{2H/h}(Z_{1H/h} + Z_{2H/h} \tan\beta)}{
 g_2^2 \cos^2\theta_W (1+\tan^2\beta)}
\left(1+\tan^2\theta_{\tilde \tau } \right)
\nonumber\\
& &
- \tan\theta_{\tilde \tau } \frac{\sqrt{2}
(Z_{1H/h} + Z_{2H/h} \tan\beta) }{g_2 \cos^2\theta_W M_W \sqrt{1+\tan^2\beta}}
\times
\nonumber\\
& &
\times
\left( Z_{2H/h} A_{\tau } +  Z_{1H/h} (A_{\tau}'+\mu^\star Y_{\tau })  + h.c. \right)
\; ,
\eea
where $Z$ is the matrix which diagonalises the Higgs mass
matrix, $Y_{\tau }$ is the tau Yukawa coupling, $ A_{\tau}, A_{\tau}' $ are
the SUSY breaking trilinear terms and $\mu $ the Higgs
supersymmetric mass parameter.
This contribution is suppressed  by $M_W^2/s $ for large $s$.
We can include it easily into the 4-vertex contribution by
substituting
\beq
1 \rightarrow 1 + \frac{C_H M_W^2}{ s - M_H^2 }
+ \frac{C_h M_W^2}{ s - M_h^2 } = 1 + K_H(s)\; .
\eeq
}
\end{itemize}

Now we can write the t-channel as the sum of a symmetric and
antisymmetric part, adding and subtracting a fictitious
u-channel, as
\bea
\frac{(2 p_1-p_3)^{\mu} (2p_2-p_4)^\nu}{ t-m_{\tilde \nu}^2}
\!\!\! &=& \!\!\!
\frac{1}{2} \left[
\frac{(2 p_1-p_3)^{\mu} (2p_2-p_4)^\nu}{ t-m_{\tilde \nu}^2}
+ \frac{(2 p_1-p_4)^{\nu} (2p_2-p_3)^\mu}{ u-m_{\tilde \nu}^2}
\right] \nonumber\\
&+& \frac{1}{2} \left[
\frac{(2 p_1-p_3)^{\mu} (2p_2-p_4)^\nu}{ t-m_{\tilde \nu}^2}
- \frac{(2 p_1-p_4)^{\nu} (2p_2-p_3)^\mu}{ u-m_{\tilde \nu}^2}
\right] \nonumber\\
\eea
so that we can make contact with the previous computation
and find for the symmetric and antisymmetric amplitudes
respectively:
\bea
{\cal A}_{sym}^{\mu\nu} &=&
+ i \frac{g_2^2}{4} \cos^2\theta_{\tilde \tau}
\left[\frac{(2 p_1-p_3)^{\mu} (2p_2-p_4)^{\nu}}
{t-m_{\tilde \nu}^2}
\right.\nonumber \\
& & \left. +
\frac{(2 p_1-p_4)^{\nu} (2p_2-p_3)^{\mu}}{ u-m_{\tilde \nu}^2}
+ 2 g^{\mu\nu} (1+K_H(s)) \right]
\eea
and
\bea
{\cal A}_{asym}^{\mu\nu} &=&
i \frac{g_2^2}{4} \cos^2\theta_{\tilde \tau}
\left[ \frac{(2 p_1-p_3)^{\mu} (2p_2-p_4)^{\nu}}
{t-m_{\tilde \nu}^2}
\right.\\
& & \left.
- \frac{(2 p_1-p_4)^{\nu} (2p_2-p_3)^{\mu}}{ u-m_{\tilde \nu}^2}
+ 2 \left( 1- G_Z(s) \right) \times
\right.\nonumber\\
& & \left.
\times \frac{ g^{\mu\nu} (t-u) - (2p_4+p_3)^{\mu} (p_1-p_2)^{\nu} +
(p_1-p_2)^{\mu} (2p_3+p_4)^{\nu} }{ s - M_Z^2}
\right] .
\nonumber
\eea
where $G_Z (s)=\frac{4}{3}\frac{\sin^2\theta_W}{\cos^2\theta_{\tilde \tau}}
\frac{M_Z^2}{s} $ vanishes in the limit of zero Z mass.
This coincides with the previous result for $K_H, G_Z, M_Z = 0$,
a part for a sign, which just corresponds in exchanging
$i\leftrightarrow j $.

\subsection{Polarisation sum}

The sum over the W polarisation in this case is given
by the polarisation tensor
\beq
\Pi^{\mu\mu'} = - g^{\mu\mu'} + \frac{p_3^\mu p_3^{\mu'}}{M_W^2}
\eeq
where $p_3 $ is the gauge boson momentum.

We have then for the matrix element
\beq
|{\cal M}|^2 = {\cal A}_{\mu\nu}^* {\cal A}^{\mu\nu}
- \frac{ |p_3^\mu A_{\mu\nu}|^2}{M_W^2}
- \frac{ |p_4^\nu A_{\mu\nu}|^2}{M_W^2}
+ \frac{ |p_3^\mu p_4^\nu A_{\mu\nu}|^2}{M_W^4}\; ;
\eeq
in this case neither amplitude vanishes
when contracted with the gauge boson's momentum.
Note that the second and third contributions are
related again by the symmetry $ p_3 \leftrightarrow p_4;
\nu \leftrightarrow \mu $ and are equal since the
final state has two particle with the same mass.

\subsection{Symmetric part}

We must compute the four contributions, and we have then
\begin{eqnarray}
{\cal A}_{\mu\nu}^* {\cal A}^{\mu\nu}
&=&
\frac{g_2^4 \cos^4\theta_{\tilde \tau}}{2}
\left[
1 +
 \frac{(m_{\tilde \nu}^2+m_{\tilde \tau}^2-M_W^2/2 )^2}{(t-m_{\tilde \nu}^2)^2}
+ 2 K_H(s) (1+K_H(s))
\right. \\
& & \left.
+ \frac{1}{2} \frac{1}{t-m_{\tilde \nu}^2}
\left( 4 (m_{\tilde \nu}^2+m_{\tilde \tau}^2)- 2 M_W^2
+ s -  4 (m_{\tilde \nu}^2+m_{\tilde \tau}^2) + 2 M_W^2
\right.\right.  \nonumber\\
& & \left. \left.
- \frac{(s - 4 m_{\tilde \tau}^2 + M_W^2)^2}{s + 2 (m_{\tilde \nu}^2-m_{\tilde \tau}^2- M_W^2)}
+ K_H(s) (s - 4 m_{\tilde \nu}^2- 4 m_{\tilde \tau}^2 + 2 M_W^2 ) \right)
\right] \nonumber\\
&=&
\frac{g_2^4 \cos^4\theta_{\tilde \tau}}{2}
\left[
1+\frac{(m_{\tilde \nu}^2+m_{\tilde \tau}^2-M_W^2/2 )^2}{(t-m_{\tilde \nu}^2)^2}
+ 2 K_H(s) (1+K_H(s))
\right. \\
& & \left.
+ \frac{1}{t-m_{\tilde \nu}^2}
\left(  m_{\tilde \nu}^2+ 3 m_{\tilde \tau}^2 - 2 M_W^2
- \frac{1}{2} \frac{(2 m_{\tilde \tau}^2 +2 m_{\tilde \nu}^2- 3 M_W^2)^2}
{s + 2 (m_{\tilde \nu}^2-m_{\tilde \tau}^2- M_W^2)}
\right.\right. \nonumber\\
& & \left. \left.
+\frac{K_H(s)}{2} ( s - 4 m_{\tilde \nu}^2- 4 m_{\tilde \tau}^2 + 2 M_W^2 )
\right)
\right] \nonumber\; ,
\end{eqnarray}
which in the limit of vanishing $M_W $ and
$ m_{\tilde \nu} =  m_{\tilde \tau} $
coincides with our old result.

The other pieces give instead
\begin{eqnarray}
\frac{ |p_3^\mu A_{\mu\nu}|^2}{M_W^2}
+ \frac{ |p_4^\nu A_{\mu\nu}|^2}{M_W^2}
&= & \frac{g_2^4 \cos^4\theta_{\tilde \tau}}{4}
\left[
\frac{(m_{\tilde \nu}^2 - m_{\tilde \tau}^2)^2}{M_W^2}
\left(
\frac{2 m_{\tilde \nu}^2+ 2 m_{\tilde \tau}^2 - M_W^2}{(t-m_{\tilde \nu}^2)^2}
\right.\right. \\
& & \left. \left.
+ \frac{2}{t-m_{\tilde \nu}^2}
\frac{2 m_{\tilde \nu}^2+ 2 m_{\tilde \tau}^2 - 3 M_W^2)^2}
{s + 2 (m_{\tilde \nu}^2-m_{\tilde \tau}^2- M_W^2)}
\right)
\right.\nonumber \\
& & \left.
-  4  K_H(s) \frac{m_{\tilde \nu}^2 - m_{\tilde \tau}^2 }{M_W^2} \left(
1 + \frac{m_{\tilde \nu}^2 - m_{\tilde \tau}^2+ s/2}{t-m_{\tilde \nu}^2}
\right)
+ 2 K_H^2 (s) \right]
\nonumber
\end{eqnarray}
and the last part:
\begin{eqnarray}
\frac{ |p_3^\mu p_4^\nu A_{\mu\nu}|^2}{M_W^4}
& =&
\frac{g_2^4 \cos^4\theta_{\tilde \tau}}{4} \left[
\frac{ (m_{\tilde \nu}^2 - m_{\tilde \tau}^2)^2}{M_W^4}
\left(
1 + \frac{1}{2} \frac{(m_{\tilde \nu}^2-m_{\tilde \tau}^2)^2}{(t - m_{\tilde \nu}^2)^2}
\right.\right. \\
& & \left. \left.
+ 2 \frac{m_{\tilde \nu}^2-m_{\tilde \tau}^2}{t - m_{\tilde \nu}^2}
\left(
1 - \frac{1}{4}  \frac{m_{\tilde \nu}^2-m_{\tilde \tau}^2}{s + 2(m_{\tilde \nu}^2-m_{\tilde \tau}^2 + M_W^2)}
\right) \right)
\right.\nonumber \\
& & \left.
+ K_H(s) \frac{m_{\tilde \nu}^2 - m_{\tilde \tau}^2}{M_W^2}
\left( \frac{s}{M_W^2} - 2 \right) \left(
1 + \frac{m_{\tilde \nu}^2-m_{\tilde \tau}^2}{t - m_{\tilde \nu}^2}
\right)
\right.\nonumber \\
& & \left.
+ \frac{1}{4} K_H^2 (s)  \left( \frac{s}{M_W^2} - 2 \right)^2
\right]
\nonumber \; .
\end{eqnarray}
Both these contributions vanish in the limit of equal
stau and sneutrino masses and zero gauge boson mass as they should.

So summing all together the result is
\begin{eqnarray}
|{\cal M}_{sym}|^2 &=&
\frac{g_2^4 \cos^4\theta_{\tilde \tau}}{2}
\left[
1+\frac{(m_{\tilde \nu}^2+m_{\tilde \tau}^2-M_W^2/2 )^2}{(t-m_{\tilde \nu}^2)^2}
\right. \\
& & \left.
+ \frac{1}{t-m_{\tilde \nu}^2}
\left(  m_{\tilde \nu}^2+ 3 m_{\tilde \tau}^2 - 2 M_W^2
- \frac{1}{2} \frac{(2 m_{\tilde \tau}^2 +2 m_{\tilde \nu}^2- 3 M_W^2)^2}
{s + 2 (m_{\tilde \nu}^2-m_{\tilde \tau}^2- M_W^2)} \right)
\right.  \nonumber\\
& & \left.
- \frac{ (m_{\tilde \nu}^2 - m_{\tilde \tau}^2)^2}{M_W^2 (t-m_{\tilde \nu}^2)}
\left(
\frac{ m_{\tilde \nu}^2+ m_{\tilde \tau}^2 -M_W^2/2 }{t-m_{\tilde \nu}^2}
+ \frac{ 2 m_{\tilde \tau}^2 +2 m_{\tilde \nu}^2- 3 M_W^2}
{s + 2 (m_{\tilde \nu}^2-m_{\tilde \tau}^2- M_W^2)}
\right)\right. \nonumber\\
& & \left.
+ \frac{1}{2}
\frac{ (m_{\tilde \nu}^2 - m_{\tilde \tau}^2)^2}{M_W^4}
\left(
1 + \frac{1}{2} \frac{(m_{\tilde \nu}^2-m_{\tilde \tau}^2)^2}{(t - m_{\tilde \nu}^2)^2}
\right.\right. \nonumber \\
& & \left. \left.
+ 2 \frac{m_{\tilde \nu}^2-m_{\tilde \tau}^2}{t - m_{\tilde \nu}^2}
\left(
1 - \frac{1}{4}  \frac{m_{\tilde \nu}^2-m_{\tilde \tau}^2}{s + 2(m_{\tilde \nu}^2-m_{\tilde \tau}^2 + M_W^2)}
\right) \right)
\right.\nonumber \\
& & \left.
+ K_H(s) \left( 2 - 3 \frac{ m_{\tilde \nu}^2 - m_{\tilde \tau}^2}{M_W^2}
+ \frac{s}{2 M_W^2} \frac{ m_{\tilde \nu}^2 - m_{\tilde \tau}^2}{M_W^2} \right)
\right. \nonumber\\
& & \left.
+ \frac{K_H(s)}{t-m_{\tilde \nu}^2}
\left(
\frac{s}{2} - 2 m_{\tilde \nu}^2 - 2 m_{\tilde \tau}^2 + M_W^2
\right.\right. \nonumber\\
& & \left.\left.
- \frac{ m_{\tilde \nu}^2 - m_{\tilde \tau}^2}{M_W^2}
\left( s +  3 (m_{\tilde \nu}^2 - m_{\tilde \tau}^2) ( 1 -
\frac{s}{2M_W^2} )
\right)
\right)
\right. \nonumber\\
& & \left.
+ K_H^2(s) \left( \frac{7}{2} - \frac{s}{2 M_W^2} + \frac{s^2}{8 M_W^4}
\right)
\right]\; . \nonumber
\end{eqnarray}
Note that the in the limit of large $s$, $ s K_H(s) $ remains finite and therefore there
is no problem with unitarity.

\subsection{Antisymmetric part}

The antisymmetric piece is more involved. We have
\begin{eqnarray}
{\cal A}_{\mu\nu}^* {\cal A}^{\mu\nu}
&=&
\frac{g_2^4 \cos^4\theta_{\tilde \tau}}{4}
\left[
2 + 2
\frac{(m_{\tilde \nu}^2+m_{\tilde \tau}^2-M_W^2/2 )^2}{(t-m_{\tilde \nu}^2)^2}
\right.\\
& & \left.
+ \frac{1}{t-m_{\tilde \nu}^2}
\left( 4 (m_{\tilde \nu}^2+m_{\tilde \tau}^2) - 2 M_W^2
+
\frac{(s - 4 m_{\tilde \tau}^2 + M_W^2)^2}{s + 2 (m_{\tilde \nu}^2-m_{\tilde \tau}^2- M_W^2)} \right)
\right.\nonumber\\
& & \left.
+ \left(1- G_Z (s)\right)^2
\frac{5/2 (t-u)^2 - 4(s- 4 m_{\tilde \tau}^2 )(s+M_W^2/2)}{(s-M_Z^2)^2}
\right.\nonumber\\
& & \left.
- \left(1- G_Z(s) \right)
\frac{ 2 (t-u) + 4(s-4 m_{\tilde \tau}^2)}{s-M_Z^2}
\right.\nonumber\\
& & \left.
+  \left(1- G_Z(s) \right)
\frac{ (t-u) (3 s -2 m_{\tilde \nu}^2 - 6 m_{\tilde \tau}^2 + 2 M_W^2)}{(s-M_Z^2) (t-m_{\tilde \nu}^2)}
\right.\nonumber\\
& & \left.
- \left(1- G_Z(s) \right)
\frac{ 4 (s+m_{\tilde \nu}^2 - m_{\tilde \tau}^2 )
(s-4 m_{\tilde \tau}^2)}{(s-M_Z^2) (t-m_{\tilde \nu}^2)}
\right] \nonumber
\end{eqnarray}
which in the limit of vanishing $M_W, M_Z $ and
$ m_{\tilde \nu} =  m_{\tilde \tau} $
coincides with our old result.

The other pieces give instead
\begin{eqnarray}
& &
\frac{ |p_3^\mu A_{\mu\nu}|^2}{M_W^2}
+ \frac{ |p_4^\nu A_{\mu\nu}|^2}{M_W^2} =
\frac{g_2^4 \cos^4\theta_{\tilde \tau}}{2} \times
\\
&\times &
\left[
\frac{1}{2} \frac{(t-u)^2}{(s-M_Z^2)^2}
\left(1- G_Z (s) \right)
\left( \frac{M_Z^2}{M_W^2} -\frac{1}{2}
- G_Z(s) \left( \frac{s}{M_W^2} - \frac{1}{2}\right) \right)
\right. \nonumber
\\
& & \left.
-  \frac{(s - 4 m_{\tilde \tau}^2) M_W^2}{(s-M_Z^2)^2}
\left( \frac{M_Z^2}{M_W^2} -1 - G_Z(s) ( \frac{s}{M_W^2} -1) \right)^2
\right. \nonumber
\\
& & \left.
- \frac{m_{\tilde \nu}^2 - m_{\tilde \tau}^2}{s-M_Z^2}
\frac{ t - u - 2 (s -  4 m_{\tilde \tau}^2)}{t-m_{\tilde \nu}^2}
\left( \frac{M_Z^2}{M_W^2} -1 - G_Z(s) ( \frac{s}{M_W^2} -1) \right)
\right. \nonumber
\\
& & \left.
- \frac{ (m_{\tilde \nu}^2 - m_{\tilde \tau}^2)^2}{M_W^2 (s-M_Z^2)}
\left(1 - G_Z(s) \right) \frac{t- u}{t - m_{\tilde \nu}^2}
\right. \nonumber\\
& & \left.
+ \frac{(m_{\tilde \nu}^2-m_{\tilde \tau}^2)^2}{M_W^2 ( t-m_{\tilde \nu}^2)}
\left( 1 + \frac{ s - 4 m_{\tilde \tau}^2 + M_W^2}{
s + 2 (m_{\tilde \nu}^2 - m_{\tilde \tau}^2 - M_W^2)}
+ \frac{m_{\tilde \nu}^2+ m_{\tilde \tau}^2 - M_W/2}{t-m_{\tilde \nu}^2}
\right) \right]\; ; \nonumber
\end{eqnarray}
in the limit of vanishing $ m_{\tilde \nu}^2 - m_{\tilde \tau}^2,
M_Z, M_W $ masses keeping $ M_Z/M_W \rightarrow 1 $ we have
\begin{eqnarray}
\frac{ |p_3^\mu A_{\mu\nu}|^2}{M_W^2}
+ \frac{ |p_4^\nu A_{\mu\nu}|^2}{M_W^2} \rightarrow
\frac{g_2^4 \cos^4\theta_{\tilde \tau}}{4}
\frac{(t-u)^2}{2 s} \; ,
\end{eqnarray}
as expected from the QCD result.

The last part gives instead
\begin{eqnarray}
\frac{ |p_3^\mu p_4^\nu A_{\mu\nu}|^2}{M_W^4}
& =&
\frac{g_2^4 \cos^4\theta_{\tilde \tau}}{16 M_W^4}
\left[
\frac{(t-u)^2}{(s-M_Z^2)^2}
\left(M_Z^2 - G_Z(s) s \right)^2
\right.\\
& & \left.
- 4 (m_{\tilde \nu}^2 - m_{\tilde \tau}^2 )^2
\frac{t-u}{(t- m_{\tilde \nu}^2)( s-M_Z^2)}
\left( M_Z^2 - G_Z(s) s \right)
\right.\nonumber\\
& & \left.
+ 2 \frac{ (m_{\tilde \nu}^2 - m_{\tilde \tau}^2)^4}{(t-m_{\tilde \nu}^2)}
\left( \frac{1}{(t-m_{\tilde \nu}^2)}+ 2
\frac{1}{s
+ 2( m_{\tilde \nu}^2-m_{\tilde \tau}^2-M_W^2)}\right)
\right]\; . \nonumber
\end{eqnarray}
Note that this contribution does not vanish in the limit of equal
stop and sbottom masses and massless gauge bosons.In fact
keeping $ M_Z/M_W \rightarrow 1 $, we have
\begin{eqnarray}
\frac{ |p_3^\mu p_4^\nu A_{\mu\nu}|^2}{M_W^4}
\rightarrow
\frac{g_2^4 \cos^4\theta_{\tilde \tau}}{8}
\frac{(t-u)^2}{2 s} \; ,
\end{eqnarray}
which gives the annihilation into the Goldstone part of the
Higgs field.

We can now put all together to give
\begin{eqnarray}
|{\cal M}_{asym}|^2 &=&
\frac{g_2^4 \cos^4\theta_{\tilde \tau}}{2}
\left[
1 + \frac{(m_{\tilde \nu}^2+m_{\tilde \tau}^2-M_W^2/2 )^2}{(t-m_{\tilde \nu}^2)^2}
\right.\\
& & \left.
+ \frac{1}{t-m_{\tilde \nu}^2}
\left( 2 (m_{\tilde \nu}^2+m_{\tilde \tau}^2) -  M_W^2
+  \frac{1}{2}
\frac{(s - 4 m_{\tilde \tau}^2 + M_W^2)^2}{s + 2 (m_{\tilde \nu}^2-m_{\tilde \tau}^2- M_W^2)} \right)
\right.\nonumber\\
& & \left.
+ \frac{1}{2} \left(1- G_Z (s)\right)^2
\frac{5/2 (t-u)^2 - 4(s- 4 m_{\tilde \tau}^2 )(s+M_W^2/2)}{(s-M_Z^2)^2}
\right.\nonumber\\
& & \left.
-  \left(1- G_Z(s) \right)
\frac{ (t-u) + 2(s-4 m_{\tilde \tau}^2)}{s-M_Z^2}
\right.\nonumber\\
& & \left.
+ \frac{1}{2} \left(1- G_Z(s) \right)
\frac{ (t-u) (3 s - 2 m_{\tilde \nu}^2 - 6 m_{\tilde \tau}^2 + 2 M_W^2)}{(s-M_Z^2) (t-m_{\tilde \nu}^2)}
\right.\nonumber\\
& & \left.
-  \left(1- G_Z(s) \right)
\frac{ 2 (s+m_{\tilde \nu}^2 - m_{\tilde \tau}^2 )
(s-4 m_{\tilde \tau}^2)}{(s-M_Z^2) (t-m_{\tilde \nu}^2)}
\right.\nonumber\\
& & \left.
 - \frac{(t-u)^2}{2 (s-M_Z^2)^2}
\left(1- G_Z (s) \right)
\left( \frac{M_Z^2}{M_W^2} - \frac{1}{2}
- G_Z(s) \left( \frac{s}{M_W^2} -\frac{1}{2}\right) \right)
\right. \nonumber
\\
& & \left.
+  \frac{ M_W^2 (s - 4 m_{\tilde \tau}^2)}{(s-M_Z^2)^2}
\left( \frac{M_Z^2}{M_W^2} - 1 - G_Z(s) ( \frac{s}{M_W^2} - 1) \right)^2
\right. \nonumber
\\
& & \left.
+ \frac{m_{\tilde \nu}^2 - m_{\tilde \tau}^2}{s-M_Z^2}
\left( \frac{M_Z^2}{M_W^2} - 1 - G_Z(s) (\frac{s}{M_W^2} - 1) \right)
\frac{ t - u - 2 (s -  4 m_{\tilde \tau}^2)}{t-m_{\tilde \nu}^2}
\right. \nonumber
\\
& & \left.
+ \frac{ (m_{\tilde \nu}^2 - m_{\tilde \tau}^2)^2}{M_W^2 (s-M_Z^2)}
\left(1 - G_Z(s) \right) \frac{t- u}{t - m_{\tilde \nu}^2}
\right. \nonumber\\
& & \left.
-  \frac{(m_{\tilde \nu}^2-m_{\tilde \tau}^2)^2}{M_W^2 (t-m_{\tilde \nu}^2)^2}
\left(  m_{\tilde \nu}^2+ m_{\tilde \tau}^2 - M_W/2 \right)
\right. \nonumber\\
& & \left.
-  \frac{(m_{\tilde \nu}^2-m_{\tilde \tau}^2)^2}{M_W^2 (t-m_{\tilde \nu}^2)}
\left(
1 + \frac{ s - 4 m_{\tilde \tau}^2 + M_W^2}{
s + 2 (m_{\tilde \nu}^2 - m_{\tilde \tau}^2 - M_W^2)}
\right)
\right. \nonumber\\
& & \left.
+  \frac{(t-u)^2}{8 (s-M_Z^2)^2}
\left( \frac{M_Z^2}{M_W^2} - G_Z(s) \frac{s}{M_W^2} \right)^2
\right.\nonumber\\
& & \left.
- \frac{1}{2} \frac{(m_{\tilde \nu}^2 - m_{\tilde \tau}^2 )^2}{M_W^2}
\frac{t-u}{(t- m_{\tilde \nu}^2)( s-M_Z^2)}
\left( \frac{M_Z^2}{M_W^2} - G_Z(s) \frac{s}{M_W^2} \right)
\right.\nonumber\\
& & \left.
+ \frac{1}{4} \frac{ (m_{\tilde \nu}^2 - m_{\tilde \tau}^2)^4}{M_W^4 (t-m_{\tilde \nu}^2)}
\left( \frac{1}{(t-m_{\tilde \nu}^2)}+ 2
\frac{1}{s
+ 2( m_{\tilde \nu}^2-m_{\tilde \tau}^2-M_W^2)}\right)
\right]\; . \nonumber
\end{eqnarray}

Note that to reduce these expressions in terms of only the $t$
variable, we have used the simple decompositions, i.e.
from $ s + t+ u = 2 m_{\tilde t} ^2 + 2 M_W^2 $ one obtains
\beq
{1\over (t-m_{\tilde \nu}^2)(u-m_{\tilde \nu}^2)} =
- {1\over s + 2 (m_{\tilde \nu}^2 -m_{\tilde \tau}^2 -  M_W^2)}
\left( {1\over t-m_{\tilde \nu}^2} + {1\over u-m_{\tilde \nu}^2}\right) .
\label{utWW-identity}
\eeq

\subsection{Results for the cross section}

We can integrate the matrix element to obtain the cross section in the two cases:
\begin{eqnarray}
\sigma_{sym} (s)
&=& \frac{g_2^4 \cos^4\theta_{\tilde \tau}}{32 \pi (s - 4 m_{\tilde \tau}^2) }
\sqrt{ \left(1 - \frac{4 m_{\tilde \tau}^2}{s} \right)
 \left(1 - \frac{4 M_W^2}{s} \right)}
\left[
1 + \frac{1}{2}
\frac{ (m_{\tilde \nu}^2 - m_{\tilde \tau }^2)^2}{M_W^4}
\right. \nonumber\\
& & \left.
+ \frac{(m_{\tilde \nu}^2+m_{\tilde \tau}^2-M_W^2/2 )^2}{
m_{\tilde \nu}^2 (s + m_{\tilde \nu}^2 - 2 m_{\tilde \tau}^2- 2 M_W^2)+ (m_{\tilde \tau}^2-M_W^2 )^2 }
\times
\right. \nonumber\\
& & \left.
\times \left( 1 - \frac{ (m_{\tilde \nu}^2 - m_{\tilde \tau}^2)^2}{M_W^2 (2 m_{\tilde \nu}^2+2 m_{\tilde \tau}^2-M_W^2) } \right)^2
\right. \nonumber\\
& & \left.
+ K_H(s) \left( 2 - 3 \frac{ m_{\tilde \nu}^2 - m_{\tilde \tau}^2}{M_W^2}
+ \frac{s}{2 M_W^2} \frac{ m_{\tilde \nu}^2 - m_{\tilde \tau}^2}{M_W^2} \right)
\right. \nonumber\\
& & \left.
+ K_H^2(s) \left( \frac{7}{2} - \frac{s}{2 M_W^2} + \frac{s^2}{8 M_W^4} \right)
\right. \nonumber\\
& & \left.
+ \frac{ Ln (s)}{\sqrt{ \left(s - 4 m_{\tilde \tau}^2 \right)
 \left(s - 4 M_W^2 \right)}}
\left(
m_{\tilde \nu}^2+ 3 m_{\tilde \tau}^2 - 2 M_W^2
\right.\right.  \nonumber\\
& & \left. \left.
- \frac{1}{2} \frac{(2 m_{\tilde \tau}^2 +2 m_{\tilde \nu}^2- 3 M_W^2)^2}
{s + 2 (m_{\tilde \nu}^2-m_{\tilde \tau}^2- M_W^2)}
- \frac{ (m_{\tilde \nu}^2 - m_{\tilde \tau}^2)^2}{M_W^2}
\frac{2 m_{\tilde \tau}^2 +2 m_{\tilde \nu}^2- 3 M_W^2}
{s + 2 (m_{\tilde \nu}^2-m_{\tilde \tau}^2- M_W^2)}
\right.
\right. \nonumber\\
& & \left.\left.
+ \frac{ (m_{\tilde \nu}^2 - m_{\tilde \tau}^2)^3}{M_W^4}
\left(1 - \frac{1}{4} \frac{m_{\tilde \nu}^2 - m_{\tilde \tau}^2}{s + 2 (m_{\tilde \nu}^2-m_{\tilde \tau}^2- M_W^2)}
\right)
\right.\right.  \nonumber\\
& & \left. \left.
+ K_H(s)
\left(
\frac{s}{2} - 2 m_{\tilde \nu}^2 - 2 m_{\tilde \tau}^2 + M_W^2
- \frac{ s (m_{\tilde \nu}^2 - m_{\tilde \tau}^2)}{M_W^2}
\right. \right.
\right. \nonumber\\
& & \left.\left.\left.
-  3 \frac{ (m_{\tilde \nu}^2 - m_{\tilde \tau}^2)^2}{M_W^2}
\left( 1 - \frac{s}{2M_W^2} \right)
\right) \right)
\right]\; . \nonumber
\end{eqnarray}

where
\begin{eqnarray}
Ln (s) &=& \ln \left[
 \frac{s + 2 (m_{\tilde \nu}^2 - m_{\tilde \tau}^2- M_W^2) -  \sqrt{ \left(s - 4 m_{\tilde \tau}^2 \right)
 \left(s - 4 M_W^2 \right)}}{
 s + 2( m_{\tilde \nu}^2 - m_{\tilde \tau}^2- M_W^2) +
 \sqrt{ \left(s - 4 m_{\tilde \tau}^2 \right)
 \left(s - 4 M_W^2 \right)}}
 \right]
\end{eqnarray}

The antisymmetric part gives instead:
\begin{eqnarray}
\sigma_{asym} (s)
&=& \frac{g_2^4 \cos^4\theta_{\tilde \tau}}{32 \pi (s - 4 m_{\tilde \tau}^2) }
\sqrt{ \left(1 - \frac{4 m_{\tilde \tau}^2}{s} \right)
 \left(1 - \frac{4 M_W^2}{s} \right)} \times
\\
& &
 \times \left[ 1
+ \left(1- G_Z(s) \right)
\frac{ s - 2 m_{\tilde \nu}^2 +2  m_{\tilde \tau}^2 + 2 M_W^2}{s-M_Z^2}
\right.\nonumber\\
& & \left.
- \left(1- G_Z(s) \right)^2
\frac{ (s - 4 m_{\tilde \tau}^2)(2 s  + M_W^2)}{(s-M_Z^2)^2}
\right.\nonumber\\
& & \left.
+  \frac{ M_W^2 (s - 4 m_{\tilde \tau}^2)}{(s-M_Z^2)^2}
\left( \frac{M_Z^2}{M_W^2} - 1 - G_Z(s) ( \frac{s}{M_W^2} - 1) \right)^2
\right. \nonumber
\\
& & \left.
+ 2 \frac{m_{\tilde \nu}^2 - m_{\tilde \tau}^2}{s-M_Z^2}
\left( \frac{M_Z^2}{M_W^2} - 1 - G_Z(s) (\frac{s}{M_W^2} - 1) \right)
\right. \nonumber
\\
& & \left.
+ 2 \frac{ (m_{\tilde \nu}^2 - m_{\tilde \tau}^2)^2}{M_W^2 (s-M_Z^2)}
\left(1 - G_Z(s) \right)
\right. \nonumber\\
& & \left.
- \frac{(m_{\tilde \nu}^2 - m_{\tilde \tau}^2 )^2}{M_W^2 ( s-M_Z^2)}
\left( \frac{M_Z^2}{M_W^2} - G_Z(s) \frac{s}{M_W^2} \right)
\right.\nonumber\\
& & \left.
+ \frac{(m_{\tilde \nu}^2+m_{\tilde \tau}^2-M_W^2/2 )^2}{
m_{\tilde \nu}^2 (s + m_{\tilde \nu}^2 - 2 m_{\tilde \tau}^2- 2 M_W^2)
+ (m_{\tilde \tau}^2-M_W^2 )^2 }
\times
\right. \nonumber\\
& & \left.
\times \left( 1 -
\frac{ (m_{\tilde \nu}^2 - m_{\tilde \tau}^2)^2}{M_W^2 (2 m_{\tilde \nu}^2+2 m_{\tilde \tau}^2-M_W^2) }
\right)^2
\right. \nonumber\\
& & \left.
+ \frac{(s - 4 m_{\tilde \tau}^2)(s - 4 M_W^2) }{ 24\; (s-M_Z^2)^2}
 \left( 10  ( 1 - G_Z(s) )^2
 + \frac{M_Z^4}{M_W^4} \left(1- G_Z (s)\frac{s}{M_Z^2} \right)^2
 \right. \right. \nonumber\\
& & \left. \left.
- 4 \left(1- G_Z (s) \right)
\left( \frac{M_Z^2}{M_W^2} - \frac{1}{2}
- G_Z(s) \left( \frac{s}{M_W^2} -\frac{1}{2}\right) \right)
\right)
\right. \nonumber\\
& & \left.
+ \frac{1}{2} \frac{ Ln (s)}{\sqrt{ \left(s - 4 m_{\tilde \tau}^2 \right)
 \left(s - 4 M_W^2 \right)}}
\left(
\frac{(s - 4 m_{\tilde \tau}^2 + M_W^2)^2}{s + 2 (m_{\tilde \nu}^2-m_{\tilde \tau}^2- M_W^2)}
+
4 (m_{\tilde \nu}^2+m_{\tilde \tau}^2)
\right.\right.\nonumber\\
& & \left.\left.
-  2 M_W^2  -  \left(1- G_Z(s) \right)
\frac{ s ( s  -  20 m_{\tilde \tau}^2 + 8 M_W^2)}{s-M_Z^2}
\right.\right.\nonumber\\
& & \left.\left.
-  4 \left(1- G_Z(s) \right)
\frac{
M_W^2 ( 4 m_{\tilde \tau}^2-M_W^2 ) + m_{\tilde \nu}^2 (m_{\tilde \nu}^2 - m_{\tilde \tau}^2)}{s-M_Z^2}
\right.\right.\nonumber\\
& & \left.\left.
- 2 \frac{m_{\tilde \nu}^2 - m_{\tilde \tau}^2}{s-M_Z^2}
\left( \frac{M_Z^2}{M_W^2} - 1 - G_Z(s) (\frac{s}{M_W^2} - 1) \right)
(s - 2  m_{\tilde \nu}^2- 6 m_{\tilde \tau}^2  +2 M_W^2 )
\right.\right.\nonumber\\
& & \left.\left.
+ 2 \frac{ (m_{\tilde \nu}^2 - m_{\tilde \tau}^2)^2}{M_W^2 (s-M_Z^2)}
\left(1 - G_Z(s) \right) (s + 2  m_{\tilde \nu}^2- 2 m_{\tilde \tau}^2  - 2 M_W^2 )
\right.\right. \nonumber\\
& & \left.\left.
-  2 \frac{(m_{\tilde \nu}^2-m_{\tilde \tau}^2)^2}{M_W^2}
\left(
1 + \frac{ s - 4 m_{\tilde \tau}^2 + M_W^2}{
s + 2 (m_{\tilde \nu}^2 - m_{\tilde \tau}^2 - M_W^2)}
\right)
\right.\right.  \nonumber\\
& & \left. \left.
-  \frac{(m_{\tilde \nu}^2 - m_{\tilde \tau}^2 )^2}{M_W^2}
\frac{s + 2 (m_{\tilde \nu}^2 - m_{\tilde \tau}^2 - M_W^2)   }{ s-M_Z^2}
\left( \frac{M_Z^2}{M_W^2} - G_Z(s) \frac{s}{M_W^2} \right)
\right.\right.  \nonumber\\
& & \left. \left.
+  \frac{ (m_{\tilde \nu}^2 - m_{\tilde \tau}^2)^4}{M_W^4
(s + 2( m_{\tilde \nu}^2-m_{\tilde \tau}^2-M_W^2))}
\right) \right]\; .
 \nonumber
\end{eqnarray}



\begin{thebibliography}{99}
\addcontentsline{toc}{section}{References}

\def\apj#1#2#3{{\it Astrophys. J. }{\bf #1} (#2) #3}
\def\ibid#1#2#3{{\it ibid\/} {\bf #1} (#2) #3}

\bibitem{ch-relics}
For a compilation of bounds on charged or coloured
relics, see the Particle Data Group,
W.-M. Yao et al., J. Phys. G 33, 1 (2006)

\bibitem{axinolsp}
L.~Covi, J.~E.~Kim and L.~Roszkowski,
  Phys.\ Rev.\ Lett.\  {\bf 82} (1999) 4180
  [arXiv:hep-ph/9905212];
L.~Covi, H.~B.~Kim, J.~E.~Kim and L.~Roszkowski,
  JHEP {\bf 0105} (2001) 033
  [arXiv:hep-ph/0101009]; 
 L.~Covi, L.~Roszkowski, R.~Ruiz de Austri and M.~Small,
  JHEP {\bf 0406} (2004) 003
  [arXiv:hep-ph/0402240].

\bibitem{crs02}
L.~Covi, L.~Roszkowski and M.~Small,
  JHEP {\bf 0207} (2002) 023
  [arXiv:hep-ph/0206119].

\bibitem{axinolsp2}
A.~Brandenburg and F.~D.~Steffen,
  JCAP {\bf 0408} (2004) 008
  [arXiv:hep-ph/0405158].
K.~Y.~Choi, L.~Roszkowski and R.~Ruiz de Austri,
  JHEP {\bf 0804} (2008) 016
  [arXiv:0710.3349 [hep-ph]].
H.~Baer and H.~Summy,
  arXiv:0803.0510 [hep-ph].


\bibitem{gravitinolsp}
J.~L.~Feng, A.~Rajaraman and F.~Takayama,
  Phys.\ Rev.\ Lett.\  {\bf 91} (2003) 011302
  [arXiv:hep-ph/0302215];
J.~R.~Ellis, K.~A.~Olive, Y.~Santoso and V.~C.~Spanos,
  Phys.\ Lett.\  B {\bf 588} (2004) 7
  [arXiv:hep-ph/0312262];
J.~L.~Feng, S.~f.~Su and F.~Takayama,
  Phys.\ Rev.\  D {\bf 70} (2004) 063514
  [arXiv:hep-ph/0404198];
D.~G.~Cerdeno, K.~Y.~Choi, K.~Jedamzik, L.~Roszkowski and R.~Ruiz de Austri,
  JCAP {\bf 0606} (2006) 005
  [arXiv:hep-ph/0509275].
 J.~L.~Feng, B.~T.~Smith and F.~Takayama,
  Phys.\ Rev.\ Lett.\  {\bf 100} (2008) 021302
  [arXiv:0709.0297 [hep-ph]].

\bibitem{neutralNLSP}
J.~L.~Feng, S.~f.~Su and F.~Takayama,
  Phys.\ Rev.\  D {\bf 70} (2004) 075019
  [arXiv:hep-ph/0404231];
L.~Roszkowski, R.~Ruiz de Austri and K.~Y.~Choi,
  JHEP {\bf 0508} (2005) 080
  [arXiv:hep-ph/0408227].

\bibitem{stauatcolliders}
W.~Buchmuller, K.~Hamaguchi, M.~Ratz and T.~Yanagida,
  Phys.\ Lett.\  B {\bf 588} (2004) 90
  [arXiv:hep-ph/0402179];
K.~Hamaguchi, Y.~Kuno, T.~Nakaya and M.~M.~Nojiri,
  Phys.\ Rev.\  D {\bf 70} (2004) 115007
  [arXiv:hep-ph/0409248];
J.~L.~Feng and B.~T.~Smith,
  Phys.\ Rev.\  D {\bf 71} (2005) 015004
  [Erratum-ibid.\  D {\bf 71} (2005) 0109904]
  [arXiv:hep-ph/0409278];
H.~U.~Martyn,
  Eur.\ Phys.\ J.\  C {\bf 48} (2006) 15
  [arXiv:hep-ph/0605257];
J.~R.~Ellis, A.~R.~Raklev and O.~K.~Oye,
  JHEP {\bf 0610} (2006) 061
  [arXiv:hep-ph/0607261];
K.~Hamaguchi, M.~M.~Nojiri and A.~de Roeck,
  JHEP {\bf 0703} (2007) 046
  [arXiv:hep-ph/0612060].


\bibitem{Fairbairn:2006gg}
  M.~Fairbairn, A.~C.~Kraan, D.~A.~Milstead, T.~Sjostrand, P.~Skands and T.~Sloan,
  Phys.\ Rept.\  {\bf 438} (2007) 1
  [arXiv:hep-ph/0611040].

\bibitem{BBNrev}
See e.g. the review by  B.~Fields and S.~Sarkar,
  in  W.~M.~Yao {\it et al.}  [Particle Data Group],
  J.\ Phys.\ G {\bf 33} (2006) 1
  [arXiv:astro-ph/0601514].

\bibitem{neutBBN}  
D.~Lindley,
  Astrophys.\ J.\  {\bf 294} (1985) 1;
M.~H.~Reno and D.~Seckel,
  Phys.\ Rev.\  D {\bf 37} (1988) 3441;
S.~Dimopoulos, R.~Esmailzadeh, L.~J.~Hall and G.~D.~Starkman,
  Astrophys.\ J.\  {\bf 330} (1988) 545;
 R.~J.~Scherrer and M.~S.~Turner,
  Astrophys.\ J.\  {\bf 331} (1988) 19
  [Astrophys.\ J.\  {\bf 331} (1988) 33];
J.~R.~Ellis, G.~B.~Gelmini, J.~L.~Lopez, D.~V.~Nanopoulos and S.~Sarkar,
  Nucl.\ Phys.\  B {\bf 373} (1992) 399.


\bibitem{CBBN}
M.~Pospelov,
  Phys.\ Rev.\ Lett.\  {\bf 98} (2007) 231301
  [arXiv:hep-ph/0605215];
K.~Kohri and F.~Takayama,
  Phys.\ Rev.\  D {\bf 76} (2007) 063507
  [arXiv:hep-ph/0605243];
M.~Kaplinghat and A.~Rajaraman,
  Phys.\ Rev.\  D {\bf 74} (2006) 103004
  [arXiv:astro-ph/0606209].


\bibitem{stauNLSP}
 R.~H.~Cyburt, J.~R.~Ellis, B.~D.~Fields, K.~A.~Olive and V.~C.~Spanos,
  JCAP {\bf 0611} (2006) 014
  [arXiv:astro-ph/0608562];
J.~Pradler and F.~D.~Steffen,
  Phys.\ Lett.\  B {\bf 648} (2007) 224
  [arXiv:hep-ph/0612291];
M.~Kawasaki, K.~Kohri and T.~Moroi,
  Phys.\ Lett.\  B {\bf 649} (2007) 436
  [arXiv:hep-ph/0703122];
J.~Pradler and F.~D.~Steffen,
  arXiv:0710.2213 [hep-ph];
J.~Kersten and K.~Schmidt-Hoberg,
  JCAP {\bf 0801} (2008) 011
  [arXiv:0710.4528 [hep-ph]];
F.~D.~Steffen,
  arXiv:0806.3266 [hep-ph].

\bibitem{wolfram}
S.~Wolfram, Phys.~Lett.~{\bf B82} (1979) 65;

\bibitem{Nardi:1990ku}
  E.~Nardi and E.~Roulet,
  Phys.\ Lett.\  B {\bf 245} (1990) 105.


\bibitem{unitarity}
K.~Griest and M.~Kamionkowski,
  Phys.\ Rev.\ Lett.\  {\bf 64} (1990) 615.


\bibitem{Cirelli07}
M.~Cirelli, A.~Strumia and M.~Tamburini,
  Nucl.\ Phys.\  B {\bf 787} (2007) 152
  [arXiv:0706.4071 [hep-ph]];

\bibitem{Sommerfeld}
A.~Sommerfeld, {\it Atombau und Spektrallinien, Band 2}, Vieweg \& Sohn (1939); \\
A.~D.~Sakharov,
  Zh.\ Eksp.\ Teor.\ Fiz.\  {\bf 18}, 631 (1948)
  [Sov.\ Phys.\ Usp.\  {\bf 34}, 375 (1991)];
J.~S.~Schwinger, {\it Particles, sources, and fields. Vol. 2,}
   Addison-Wesley (1989) (Advanced book classics series).


\bibitem{gluino-splitsusy}
  A.~Arvanitaki, C.~Davis, P.~W.~Graham, A.~Pierce and J.~G.~Wacker,
  Phys.\ Rev.\  D {\bf 72} (2005) 075011
  [arXiv:hep-ph/0504210].


\bibitem{Hisano:2006nn}
  J.~Hisano, S.~Matsumoto, M.~Nagai, O.~Saito and M.~Senami,
  Phys.\ Lett.\  B {\bf 646} (2007) 34
  [arXiv:hep-ph/0610249].


\bibitem{Freitas07}
A.~Freitas,
  Phys.\ Lett.\  B {\bf 652} (2007) 280
  [arXiv:0705.4027 [hep-ph]];

\bibitem{DMsommerfeld}
N.~Baro, F.~Boudjema and A.~Semenov,
  Phys.\ Lett.\  B {\bf 660} (2008) 550
  [arXiv:0710.1821 [hep-ph]];
J.~March-Russell, S.~M.~West, D.~Cumberbatch and D.~Hooper,
  arXiv:0801.3440 [hep-ph].


\bibitem{Strumia:2008cf}
  A.~Strumia,
  arXiv:0806.1630 [hep-ph].


\bibitem{kt91}
E.~W.~Kolb and M.~S.~Turner,
  {\it The Early Universe,}
  Front.\ Phys.\  {\bf 69} (1990) 1.

\bibitem{gg91}
P.~Gondolo and G.~Gelmini,
  Nucl.\ Phys.\  B {\bf 360} (1991) 145.

\bibitem{ahs00}
T.~Asaka, K.~Hamaguchi and K.~Suzuki,
  Phys.\ Lett.\  B {\bf 490} (2000) 136
  [arXiv:hep-ph/0005136].


\bibitem{bhsz96}
W.~Beenakker, R.~Hopker, M.~Spira and P.~M.~Zerwas,
  Nucl.\ Phys.\  B {\bf 492}, 51 (1997)
  [arXiv:hep-ph/9610490].





\bibitem{gluinoLSP}
  S.~Raby,
  Phys.\ Lett.\  B {\bf 422} (1998) 158
  [arXiv:hep-ph/9712254];
  H.~Baer, K.~m.~Cheung and J.~F.~Gunion,
  Phys.\ Rev.\  D {\bf 59} (1999) 075002
  [arXiv:hep-ph/9806361].

\bibitem{threshold}
A.~H.~Hoang,
  Phys.\ Rev.\  D {\bf 56}, 7276 (1997)
  [arXiv:hep-ph/9703404];
A.~H.~Hoang, A.~V.~Manohar, I.~W.~Stewart and T.~Teubner,
  Phys.\ Rev.\  D {\bf 65}, 014014 (2002)
  [arXiv:hep-ph/0107144].

\bibitem{Kublbeck:1992mt}
  J.~Kublbeck, H.~Eck and R.~Mertig,
  Nucl.\ Phys.\ Proc.\ Suppl.\  {\bf 29A}, 204 (1992).


\bibitem{QCDconfined}
  S.~J.~J.~Gates and O.~Lebedev,
  Phys.\ Lett.\  B {\bf 477} (2000) 216
  [arXiv:hep-ph/9912362].


\bibitem{Kang:2006yd}
  J.~Kang, M.~A.~Luty and S.~Nasri,
  arXiv:hep-ph/0611322.


\bibitem{WMAP5}
J.~Dunkley {\it et al.}  [WMAP Collaboration],
  arXiv:0803.0586 [astro-ph].

\bibitem{Lamon:2005jc}
  See R.~Lamon and R.~Durrer,
  Phys.\ Rev.\  D {\bf 73} (2006) 023507
  [arXiv:hep-ph/0506229] and references therein.

\bibitem{photon-flux}
  See K.~Abazajian, G.~M.~Fuller and W.~H.~Tucker,
  Astrophys.\ J.\  {\bf 562} (2001) 593
  [arXiv:astro-ph/0106002];
  A.~Boyarsky, A.~Neronov, O.~Ruchayskiy and M.~Shaposhnikov,
  Mon.\ Not.\ Roy.\ Astron.\ Soc.\  {\bf 370} (2006) 213
  [arXiv:astro-ph/0512509];
  G.~Bertone, W.~Buchmuller, L.~Covi and A.~Ibarra,
  JCAP {\bf 0711} (2007) 003
  [arXiv:0709.2299 [astro-ph]];
  H.~Yuksel and M.~D.~Kistler,
  Phys.\ Rev.\  D {\bf 78} (2008) 023502
  [arXiv:0711.2906 [astro-ph]] and references therein.


\bibitem{Yamagata:1993jq}
  T.~Yamagata, Y.~Takamori and H.~Utsunomiya,
  Phys.\ Rev.\  D {\bf 47} (1993) 1231.

\bibitem{Smith:1982qu}
  P.~F.~Smith, J.~R.~J.~Bennett, G.~J.~Homer, J.~D.~Lewin, H.~E.~Walford and W.~A.~Smith,
  Nucl.\ Phys.\  B {\bf 206} (1982) 333.

\bibitem{Hemmick:1989ns}
  T.~K.~Hemmick {\it et al.},
  Phys.\ Rev.\  D {\bf 41} (1990) 2074.

\bibitem{Verkerk:1991jf}
  P.~Verkerk, G.~Grynberg, B.~Pichard, M.~Spiro, S.~Zylberajch, M.~E.~Goldberg and P.~Fayet,
  Phys.\ Rev.\ Lett.\  {\bf 68} (1992) 1116.

\bibitem{Norman:1988fd}
  E.~B.~Norman, R.~B.~Chadwick, K.~T.~Lesko, R.~M.~Larimer and D.~C.~Hoffman,
  Phys.\ Rev.\  D {\bf 39} (1989) 2499.


\bibitem{BBN-kkm}
M.~Kawasaki, K.~Kohri and T.~Moroi,
  Phys.\ Rev.\  D {\bf 71} (2005) 083502
  [arXiv:astro-ph/0408426].

\bibitem{Jedamzik:2006xz}
  K.~Jedamzik,
  Phys.\ Rev.\  D {\bf 74} (2006) 103509
  [arXiv:hep-ph/0604251].

\bibitem{Kawasaki:2008qe}
  M.~Kawasaki, K.~Kohri and T.~Moroi,
  Phys.\ Lett.\  B {\bf 649} (2007) 436
  [arXiv:hep-ph/0703122];
  M.~Kawasaki, K.~Kohri, T.~Moroi and A.~Yotsuyanagi,
  arXiv:0804.3745 [hep-ph].

\bibitem{Li6}
 C.~Bird, K.~Koopmans and M.~Pospelov,
  arXiv:hep-ph/0703096;
  T.~Jittoh, K.~Kohri, M.~Koike, J.~Sato, T.~Shimomura and M.~Yamanaka,
  Phys.\ Rev.\  D {\bf 76} (2007) 125023
  [arXiv:0704.2914 [hep-ph]];
 K.~Jedamzik,
  Phys.\ Rev.\  D {\bf 77} (2008) 063524
  [arXiv:0707.2070 [astro-ph]].
 T.~Jittoh, K.~Kohri, M.~Koike, J.~Sato, T.~Shimomura and M.~Yamanaka,
  arXiv:0805.3389 [hep-ph];
 D.~Cumberbatch, K.~Ichikawa, M.~Kawasaki, K.~Kohri, J.~Silk and G.~D.~Starkman,
  Phys.\ Rev.\  D {\bf 76} (2007) 123005
  [arXiv:0708.0095 [astro-ph]];
 M.~Kusakabe, T.~Kajino, R.~N.~Boyd, T.~Yoshida and G.~J.~Mathews,
  Phys.\ Rev.\  D {\bf 76} (2007) 121302
  [arXiv:0711.3854 [astro-ph]].



\bibitem{Jedamzik:2007qk}
  K.~Jedamzik,
  JCAP {\bf 0803} (2008) 008
  [arXiv:0710.5153 [hep-ph]].


\bibitem{Hamaguchi:2007mp}
  K.~Hamaguchi, T.~Hatsuda, M.~Kamimura, Y.~Kino and T.~T.~Yanagida,
  Phys.\ Lett.\  B {\bf 650} (2007) 268
  [arXiv:hep-ph/0702274].

\bibitem{Pradler:2007is}
  J.~Pradler and F.~D.~Steffen,
  arXiv:0710.2213 [hep-ph].


\bibitem{micro}
G.~Belanger, F.~Boudjema, A.~Pukhov and A.~Semenov,
  Comput.\ Phys.\ Commun.\  {\bf 149} (2002) 103
  [arXiv:hep-ph/0112278], 
  Comput.\ Phys.\ Commun.\  {\bf 174} (2006) 577
  [arXiv:hep-ph/0405253], 
  Comput.\ Phys.\ Commun.\  {\bf 176} (2007) 367
  [arXiv:hep-ph/0607059].


\bibitem{frank06}
  F.~D.~Steffen,
  JCAP {\bf 0609} (2006) 001
  [arXiv:hep-ph/0605306].

\bibitem{bchrs05}
A.~Brandenburg, L.~Covi, K.~Hamaguchi, L.~Roszkowski and F.~D.~Steffen,
  Phys.\ Lett.\  B {\bf 617} (2005) 99
  [arXiv:hep-ph/0501287].

\bibitem{LEP-stau}
LEPSUSYWG, ALEPH, DELPHI, L3 and OPAL experiments, note LEPSUSYWG/02-05.1
(http://lepsusy.web.cern.ch/lepsusy/Welcome.html).

\bibitem{stop-gravitino}
  J.~L.~Diaz-Cruz, J.~R.~Ellis, K.~A.~Olive and Y.~Santoso,
  JHEP {\bf 0705} (2007) 003
  [arXiv:hep-ph/0701229].

\bibitem{TeVatron-stop}
J.~Nachtman, talk at the symposium ``The Hunt for Dark Matter'', Fermilab, 10--12 May 2007.

\bibitem{R-parity}
F.~Takayama and M.~Yamaguchi,
  Phys.\ Lett.\  B {\bf 485} (2000) 388
  [arXiv:hep-ph/0005214];
  W.~Buchmuller, L.~Covi, K.~Hamaguchi, A.~Ibarra and T.~Yanagida,
  JHEP {\bf 0703} (2007) 037
  [arXiv:hep-ph/0702184].




\end{thebibliography}
\end{document}